\documentclass[%
 reprint,
 amsmath,amssymb,
 aps,
endfloat
]{revtex4-2}
\usepackage{hyperref}
\usepackage{graphicx}% Include figure files
\usepackage{dcolumn}% Align table columns on decimal point
\usepackage{comment}
\usepackage[english]{babel}
\usepackage{bm}% bold math
\usepackage[labelformat=empty,font=small, justification=raggedright,format=plain]{caption}

\begin{document}

\preprint{Excitable Mechanics}

\title{Excitable mechanics embodied in a walking cilium\\
\small{Part 1 (of 3): Organelle scale}}
%\small{Part 1 (of 3) reporting the stable yet sensitive collective dynamics of ciliary flocking across scales.}}% Force line breaks with \\
%\thanks{Part 1 of 3 for a series reporting the stable yet sensitive collective dynamics of ciliary flocking across scales.}%

\author{Matthew S. Bull}
\email{bullm@stanford.edu}
\affiliation{Department of Applied Physics}

\author{Laurel A. Kroo}%
\affiliation{%
Department of Mechanical Engineering
}%

\author{Manu Prakash}
 \email{manup@stanford.edu}
\affiliation{
 Department of Bioengineering\\
 Stanford University, Stanford, CA 94305, USA
}%

\date{\today}% It is always \today, today,

\begin{abstract}
\textbf{Abstract:} Rapid transduction of sensory stimulation to action is essential for an animal to survive. To this end, most animals use the sub-second excitable and multistable dynamics of a neuromuscular system. Here, studying an animal without neurons or muscles, we report analogous excitable and multistable dynamics embedded in the physics of a 'walking' cilium. We define a 'walking cilium' as a phenomena where the locomotive force is generated through contact with a substrate which is periodically reset by steps in the direction of motion. We begin by showing that cilia can walk without specialized gait control and identify the characteristic scales of spatio-temporal height fluctuations of the tissue. With the addition of surface interaction, we construct a low-order dynamical model of this single-cilium sub-unit. In addition to studying the dynamics of a single cilium in isolation, we subsequently examine collections of these walking cilia.  En route to an emergent model for ciliary walking, we demonstrate the limits of substrate mediated synchronization between cilia. In the desynchronized limit, our model shows evidence of localized multi-stability mediated by the crosstalk between locomotive forcing and height. The out-of-equilibrium mechanics -- which govern this emergent excitable bistability -- directly control the locomotive forcing of a walking cilia. This direct coupling bypasses the role of the synaptic junctions between neurons and muscles.  We show a minimal mechanism -- trigger waves -- by which these walking cells may work together to achieve organism-scale collaboration, such as coordination of hunting strikes across $10^5$ cells without central control.

\textbf{Significance:} Eukaryotic cilia are both sensors and actuators. There is a growing field of research dedicated to understanding how stimulus is transformed into action through biochemical queues within these intricate sub-cellular structures. Here, we complement this ongoing research program by considering the hypothesis that these biochemical dynamics could be outsourced to the fast timescale response of the mechanics of an oscillating cilium. To accomplish this we studied how an emerging model organism with no muscles or neurons in the phylum placozoa (plate animals) move around using ciliary "walking". To our surprise, we found that ciliary walking can be effectively described as an excitable dynamical system. We expect that these results will enrich the discussion of how cilia can be used to coordinate the behavior of many cells without a governing neuromuscular system. 

%\begin{description}
%\item[Keywords]

%\item[A note to readers]
\textbf{A note to readers:} This work is the first of three complementary stories we have posted together describing multiple scales of ciliary flocking. While we encourage readers to read all three for a more complete picture, each of these works are written to be as self-contained as possible. In the rare case where we evoke a result from another manuscript, we make effort to motivate it as a falsifiable assumption on the grounds of existing literature and direct readers to the relevant manuscript for a more in-depth understanding.

%\item [Part 1 ]
\textbf{Part1: Excitable mechanics embodied in a walking cilium} (this work) describes the emergent mechanics of ciliary walking as a transition between ciliary swimming and ciliary stalling with increasing adhesion. 

%\item [Part 2 ] 
\textbf{Part2: Non-neuromuscular collective agility by harnessing instability in ciliary flocking} \cite{bullpart2} reports the implications of an effective rotational degree-of-freedom governing the direction of ciliary walking for agile locomotive behavior in an animal without a brain. 

%\item [Part 3 ] 
\textbf{Part3: Mobile defects born from an energy cascade shape the locomotive behavior of a headless animal}\cite{bullpart3} describes the emergent locomotive behavior of the animal in terms of a low dimensional manifold which is identified by both top-down and bottom-up approaches to find agreement.
%\end{description}
\end{abstract}

\maketitle

%\tableofcontents
\section*{Introduction}

The neuromuscular system rapidly and efficiently converts sensory stimulus into behavioral response through electro-chemically mediated action potentials \cite{sterling_principles_2015}. These signals pass between groups of cells specialized for sensing, processing, and mechanical action. This separation between sensing and actuation has encouraged thinking about an animal through its distinct components: sensory elements, processors, and actuators\cite{pfeifer2007self, Keijzer2015MovingOrganization, sterling_principles_2015}. Over the past two decades, research into embodied computation has relaxed this conceptual distinction by showing that computational tasks can be outsourced to the dynamics \cite{arcas_what_2000} and configuration of the compliant body \cite{pfeifer2007self, sterling_principles_2015}, generating significant excitement about the role of soft-materials in robotics \cite{laschi_soft_2014}. In organismic biology - this conceptual foundation has had important implications on our understanding of animal behavior\cite{weber_wing_2021, wang_neuron_2020,pratt_neural_2017}.

One of the recurring themes within these compliant physical reservoirs \cite{tanaka_recent_2019} of sufficiently nonlinear dynamics, is the emergence of the dynamical motif of excitability. Excitability and its related cousin -  bi-stability -  with a slow recovery variable is best known in the Hodgkin-Huxley equation for neurons but manifests itself much more broadly from cortical excitability \cite{Armon2018Ultra-fastHypothesis, armon_epithelial_2020, nishikawa_controlling_2017, boocock_theory_2021}, bio-electricity \cite{mcnamara_bioelectrical_2020}, non-neuronal excitability \cite{wan_origins_2021}, and hydrodynamically mediated trigger waves
\cite{mathijssen_collective_2019}. 

To better understand what role dynamics plays - it is crucial to study biological systems at multiple scales.  
 Cilia are remarkable sub-cellular structures capable of a wide array of sensing and cellular actuation \cite{gilpin_multiscale_2020} and stand-in as a good example integrating multi-scale dynamics. In its simplest form, cilia can be considered as a "unit" of motility (akin to an atom), and can be arranged in complex hierarchical architecture suiting biological function. A growing body of work has demonstrated how direct cellular control of the ciliary gait or actuation sequence can be mapped through scales from molecular details through behavior to ecology\cite{hein_natural_2016, wan2018coordination, larson_unicellular_2021}. As the field appreciates more acutely the sensitivity of ciliary behavior to biophysical details, the universal aspects become increasingly refined \cite{camalet2000generic, friedrich2016hydrodynamic}. 

Utilizing both experimental and theoretical approaches, here we study how an animal "walks" with thousands to millions of cilia on a substrate. We first experimentally demonstrated unique walking gait of individual cilia in a simple non-neuromuscular animal, \textit{Trichoplax adhaerens}, in the phylum placozoa, which uses its bottom tissue (of monociliated cells) to walk on substrates. The strong interaction of these cilia with the surface (detachment forces in the nN) \cite{kreis_adhesion_2018} combined with a cilium's finite force (10's of pN) \cite{boddeker_dynamic_2020, hill2010force} encourages us to consider substrate interaction via adhesion existing paradigms such as ciliary beating under load \cite{vilfan2006hydrodynamic, klindt2016load, larson_unicellular_2021, gadelha2010nonlinear, ohmura_simple_2018} and ciliary synchronization\cite{gilpin_multiscale_2020, wan2014rhythmicity, wan2018coordination, geyer2013cell, goldstein_elastohydrodynamic_2016, liu_transitions_2018,guo_intracellular_2021, chelakkot_synchronized_2021}. The outcome of this work is an experimentally grounded model for a walking cilium which highlights a here-to-fore under-discussed aspect of excitability \cite{wan_origins_2021} of a cilia under confinement with rich mechanics.

More specifically, in this work, we put forward a simple, non-neuromuscular animal for the study of coordinated multi-cellular mechanics and report that the fundamental unit of locomotive forcing -- a mono-ciliated cell can "walk" -- exhibiting two forms of mechanical excitability. We show how the exotic mechanics of a walking cilium can act analogously to neuromuscular excitability through a joint experimental and theoretical study spanning single-cilium measurements to whole animal behavior. Our work provides a complementary yet parallel example to the growing appreciation for cellular excitability\cite{wan_origins_2021} and its role in mapping stimulus onto action in a fashion which balances stability against noise with sensitivity to signal. The primary goal of this work is to explain ciliary walking by constraining a simple set of governing principles with a minimal set of experimentally derived physical constraints. This model by construction leads to qualitative understanding of how we may re-frame the mechanics of ciliary walking across a spectrum of timescales and parameter regimes. 

We expect that this work will contribute to a growing understanding of how animals can achieve collective stability without compromising sensitivity both in the presence and absence of a neuromuscular system\cite{Keijzer2015MovingOrganization}.

\subsection*{Model organism}

\textit{Trichoplax adhaerens} is a simple animal composed of only 6 known cell types\cite{Smith2014} in the phyla placozoa. The organism has a top-bottom symmetry breaking \cite{dubuc_dorsalventral_2019}. The top tissue is characterized by fast cellular contractions\cite{armon2018ultrafast} which are thought to aid in tissue cohesion \cite{armon_epithelial_2020}. The bottom tissue is made up mixture of mechanically distinct cells \cite{prakash_motility-induced_2021} each with their own cilium \cite{Grell1974ElektronenmikroskopischePlacozoa}. 
Its behavior is punctuated by periodic transitions\cite{Ueda1999DynamicAdhaerence} which can be stimulated by the addition of neuropeptide \cite{Senatore2017NeuropeptidergicSynapses.}. While it does not have any known muscles or neurons\cite{smith2015coordinated}, a sodium action potential has been reported\cite{romanova_sodium_2020}. 

Previous work has classified the organisms locomotion as gliding \cite{Smith2014a, sugimoto_theory_2010} enabling motility on a substrate without a head or tail\cite{ smith_coherent_2019, Armon2018Ultra-fastHypothesis, prakash_motility-induced_2021}. Here, via direct imaging of entire beat form of single cilia at high speed - we demonstrate that individual cilia exhibits a "walking" gate. Utilizing experimental, numerical and analytical approaches, we explore the utility of a new framework of cilia under confinement evoking excitable and bistable mechanics. 

\subsection*{Road map}
We begin this work by exploring the walking gait of a cilium on a sticky surface in our model animal, \textit{Trichoplax adhearens}. Via developing a novel imaging framework, we first directly visualize the walking gait of individual cilia at the bottom epithelium of the organism, at high speed. Further we experimentally measure fluctuations in tissue height for an organism undergoing locomotion on a flat substrate. Next we build a theoretical framework to understand cilia as an oscillator under confinement. Theoretically, we show that both sticky surfaces and lubricated surfaces have an effective adhesion energy which, when balanced with ciliary forcing, generates a ciliary gait propelling the organism through the collective ciliary walking. Interestingly, this walking gait intimately couples the locomotive force generation both in the tangential (direction of movement ) but also normal forces modulating height of the tissue off the substrate in a fashion which is analogous to a local field potential \cite{mcnamara_bioelectrical_2020}.

To understand this observation, we begin modeling the dynamics in ciliary shape space \cite{ma2014active, klindt2016load} using an effectively low order description of a phase-amplitude-curvature oscillator consistent with the leading order contributions of interaction with the substrate. We study this model numerically to reconstruct a parameter space characterized by three qualitatively distinct behaviors: 1) sliding, 2) walking, and 3) stalled cilium. After showing that pairs of walking cilia can remain unsynchronized for a large set of initial conditions relevant to organsimal behavior, we replace the oscillator description with an emergent, phase-cycle averaged force as a function of height. We show how to reduce the number of microscopic parameters (e.g. ciliary tip force, adhesion energy, limit cycle stiffness) by expanding to second order to show the universal character of the response under the minimal assumption that the force-velocity relationship has a maximum at an intermediate height. We use the condition of balanced torque to couple the ciliary forcing and height dynamics into a self-stabilizing differential equation which exhibits fast timescale multistable behavior in height. The resulting stable states are connected by fast switching heteroclinic orbits sensitive to mechanical stimulus and feedback from the tissue. These mutlistable dynamics are present for a broad class of parameters. We show that this emergent multistability can manifest itself in fast locomotive transitions mediated by mechanical forces and show how this can be used by the organism to coordinate rapid multicellular hunting strikes across hundreds of thousands of cells.

\section*{Direct observation of Ciliary Walking}
Most cilia swim or pump fluid at low Reynold's number (Re) through a non-reciprocal beat-profile which encloses a finite volume in confirmation space. Extensive work has characterized the modifications to fluid flow near non-slip boundary conditions and have brought these results to the context of cilia \cite{boddeker_dynamic_2020, ohmura_simple_2018}. 

In this work, we report via both experimental and theoretical approach - a new limit of cilia-substrate contact. We demonstrate that in \textit{Trichoplax adhaerens}, an organism known to crawl on surfaces - cilia (a chemomechanical oscillator) when in contact against a substrate can 'walk' without specialized gait control. We begin by presenting four primary sources of experimental support derived from experiments on ciliary walking in \textit{T. Adhearens}.

Placozoa represent a valuable opportunity to experimentally study the dynamics of cilia in high-force environments dominated by surface interaction. Previous work showed that low dimensional shape space is robust against sufficiently small external loads\cite{klindt2016load}. These findings support the conceptual picture of ciliary dynamics sweeping out a manifold living in a high dimensional ($N>>1$) space which has the property of being locally stable in all directions orthogonal ($N-1$) to the local direction of active forcing.

This is consistent with the notion that a cilium does not have an internal clock. Let's consider a thought experiment: if a cilium is stopped for a period of time, $\tau$,  and then released - the instantaneous velocity after release will not depend upon the time of being stalled, $\tau$. If such an internal clock existed and the mechanical motion was a result of entertainment to this cellular clock cycle, then you would expect to see a strong dependence of the release behavior for differing time of stall (i.e. $\tau$ would become a tool to probe the phase response curve of this coupling). The results of this experiment support the notion that a cilium is at its core a chemo-mechanical oscillator which stores its phase in configuration which can be extracted by effectively low dimensionalizing the dynamic curvature components into a phase-amplitude oscillator in 'shape' space \cite{ma2014active, klindt2016load}. 

\subsection*{Visualizing walking cilia in side view}
We have designed experiments which allow us to watch the organismal locomotion from the side for a small cross section as a organism walks upon millimeter-scale, fabricated PDMS walls. These side-view records reveal two essential observations (1) cilia 'walk' or crawl through direct attachment with the substrate and (2) height fluctuations of the underlying tissue control the interaction with the substrate (Figure 1). 

High speed movies [see figure 1D-E and supplementary movies 1, 2] of these contact points show translation or slipping consistent with a limit where ciliary actuation forces in the 10's of pN range are sufficient to peel the cilia. Combined with total internal reflection microscopy showing scatter of the evanescent field [see supplementary information Supplementary figure 2], this suite of observations supports the hypothesis that the cilia is generating an effective adhesion energy mediated by the closest point between the ciliary body and the substrate.

\subsection*{Ciliary imaging, bottom view}
In an effort to explore the ciliary attachment deeper, we developed a TIRF assay to visualize the ciliary interaction with the substrate. We use cell mask orange to stain the ciliary membrane. Using an objective TIRF (Total Internal Reflection)  setup coupling a 561nm laser into a totally internally reflected mode. Imaging the red-shifted fluorescent signal shows clear excitation by the evanescent field supporting the notion that the cilia is brought into close contact with the substrate. Despite this close contact, we further observe small amounts of ciliary contact point sliding relative to the glass substrate corresponding to a sliding point of attachment. Complementary to this sliding, we also observe distinct 'steps' in which the ciliary point of contact leaves the evanescent field before returning to the plane rapidly there after (imaged at 10 Hz) in a new location advancing in the direction of organism motion.

To gather more information beyond the evanescent field, we developed a protocol for imaging ciliary beats both in fluorescent and in bright field. For florescence, we employed epi-illumination of cell-mask orange using a powerful multi-lamp light source (Sutter Lambda XL). These experiments were limited in imaging rate to 10Hz due to the rapid motion of the individual ciliary beats and organismal translation speed at these lengthscales. These movies captures a sense for the orientational dynamics, but struggle to disentangle the dynamics of the ciliary beat.

These limitations were addressed using a carefully tuned trans-illumination imaging modality which allowed the imaging to be conducted at much higher speed (Camera limited to 100 Hz at full frame and 400 Hz sub-sampled using the sCMOS Hamamatsu Orca 4 v2+). Leveraging an asymmetry in the transmission illumination imaging of spatially extended objects such as ciliary cross-sections (above plane the cilia appears dark and below plane the cilia appears bright due to light refocusing of the ciliary body in plane), we partially reconstruct the ciliary beat both in plane and as it pulls above the imaging plane. This high temporal resolution imaging coupled with the ability to infer which portions of the cilium are above and which are below enabled us to infer the dynamics of the ciliary beat through a peel and step mode. We use this imaging modality to complement the side-view dynamics of the ciliary step and find support for the cell-to-tip traveling wave on top of a large static curvature perturbed by the presence of a stiff, sticky substrate \cite{geyer_independent_2016, klindt2016load}.

\subsection*{Ciliary activity vs adhesion in a walking cilia}
Next we present a conceptual model of "walking cilia" to more deeply understand the control mechanisms that underlie the wide range of large-scale global behaviors we observe experimentally in the organism (Figure 1). When a portion of the cilium is in contact with the substrate, we observe sharp bending to maximize cilia-surface contact. In effect, the experimental data strongly suggests there is an adhesion at play between the cilium and the surface. Based on experimental observations described above, we first set up a ground-up theoretical model (or model by construction) to provide a conceptual framework, which informs how (and which) biological experimental data is studied.

Consider the cilium as a sticky elastic rod held at a fixed angle relative to an adhesive substrate. A naive energy minimization argument suggests that the configuration which minimizes the sum of the elastic energy($E_k$) and the adhesion energy ($E_a$). These two terms compete in that the adhesion energy wants the rod to lay flat on the surface ($E_a = -\epsilon_a \ell_{contact}$) which maximizes the length of the contact between the rod and surface. However, the elastic energy comprises of deviations of the elastic structure from its preferred state ($E_k = \int_{length} k\left(\psi(s,t)-\psi_o(s,t)\right)^2 ds $, where $k$ represents the stiffness of the rod to changes of preferred phase angle and is subject to the boundary conditions determined by the cell and substrate positions). For a rod that simply wants a constant curvature (given by the value of the cilium's static curvature) then the preferred shape will be a quarter circular segment. When this rod begins interacting with the sticky surface, it will bend to increase the amount of cilium-substrate length ($\ell_{contact}$). Higher order terms in the elastic energy can result in 'buckling' where the global energy is minimized by localizing the vast majority of the bending energy to a very small region of the cilium length. This story allows us to see deeper into the observations of ciliary walking by noticing two items: (i) the cilium appears to have a region which is uniformly adhesive (e.g. the effective adhesion energy is not at a single point) and (ii) the surface adhesion is sufficient to induce buckling in the cilium with local curvatures exceeding that of the native beat.

The adhesive nature of these cilia is well supported in experiments [See Supplementary information and \cite{larson_unicellular_2021, kreis_adhesion_2018}]. However, the underlying mechanism which causes this adhesion in \textit{T. Adhaerens} is still poorly understood, and a topic for future study. It is possible that this adhesion arises directly from chemio-mechanical origins ('stickyness') with associated adhesion proteins, as in Chlamydomonus\cite{kreis_adhesion_2018}, Tetraheymena \cite{wolfe_cilia_1993}, Euplotes\cite{plumper_conjugation_1995} and in metazoans such as veliger larvae \cite{romero_capture_2010} and MDCK culture \cite{ott_primary_2012}. We model ciliary adhesion via an effective adhesion energy, $E_a$, which represents the most profitable leading order model to explore.

\subsection*{Tracking ciliary cross-sections}
In an effort to track the modification of the ciliary beat frequency with substrate interaction, we developed a technique for tracking the plane of intersection of cilia to measure step frequency. This measurement is conducted on a custom built dual plane tracking microscope (Applied Scientific Instrumentation, RAMM-2x-MIM) imaging the plane directly below the tissue surface. This allowed us to directly track and follow a group of cilia while the organism moves freely on the substrate (glass slide). The individual ciliary point of plane intersection shows up as a mobile dark point which undergoes noisy periodic sweeps signaling the presence of a step. 

This tracking microscopy technique enabled us to directly measure an accessible proxy of the frequency of ciliary beating for a small number of clearly tracked ciliary cross-sections.  Our measurements reveal three interesting aspects. (i) The ciliary step frequency is strongly modulated by interaction with the substrate with our observations ranging widely from 6 Hz on an unconfined cilium, to 0.2 Hz as observed in figure 1D-E [and SI figure 3] (ii) Nearby cilia influence the step of their neighbors. (iii) The tissue dynamics couples ciliary steps.

\subsection*{Height dynamics}
Watching placozoa walk from both the side-view (Figure 1D) and bottom-view reveals the critical role of local tissue-substrate separation. In this work, we denote this separation as a local parameter \textit{height}. There are two manners with which the local height can be modified: (1) changes to the instantaneous tissue height which is a self-organized quantity and (2) very local alterations to the environmental substrate planarity which are shorter than the wavelength established by the bending energy of the tissue. It is valuable to note that \textit{Trichoplax adhaerens} is less than 20 microns thick, with both the epithelium layers combined, thus making for a very thin compliant tissue out of plane. As the organism walks, the thin sheet clearly exhibits many local fluctuations which can be directly corroborated by examples of side view imaging as the organism walks along (see figure 1D-E). Importantly, the significant difference in height has notable effects on the beat form of the ciliary gait ranging from unencumbered to compressed into a crawl. Next we explore this important and emergent feedback loop in greater detail. 

To study the role of these fluctuations more quantitatively, we developed a method of single-frame inference of tissue height through calibrated optical sharpness. This technique relies heavily on both the optical configuration of the imaging tools as well as some of the structural details of the bottom tissue. First, we collect high spatial resolution data using a 60x oil (Nikon, NA=1.43) and a 40x oil objective acting as a condenser of a 532nm super-luminescent LED with a tracking microscope (with the Z-axis disabled). We train the focus $~3\mu m$ below the bulk tissue plane which is around $\sim `10 \mu m$ above the substrate plane. Here, cilia are only visible as a single dark dot while the bulk tissue is highlighted by intricate subcellular structure only visible when directly in the very bottom of the tissue plane. In the optical plane above this intricate structure, the tissue occlude more and more refined measurements via scattering which are reflected in an increased zeroth order aberration consistent with defocus. We can use this attenuation of high spatial frequency information to infer from a single image the local height of the tissue. To do this we develop a simple calibrated map between local image sharpness and difference between the tissue bottom plane and that of the focal plane of the microscope (see fig 1F-H). By using sharpness as a proxy for height, we can directly infer these rich spatio-temporal dynamics as the organism crawls around. 

In an attempt to characterize the effective length and time-scales of these spatio-temporal dynamics, we plot up the length and time correlations. To approximately subtract off tissue displacement (and maintain the approximate map between pixel identity and tissue region), we used a Scale Invariant Feature Transformation (SIFT) to register the tissue position up to translation and rotation. Subsequently, we defined a fluctuation field by subtracting off the temporal mean of the height at each pixel, as a method for removing any systematic bias between the sample plane and the imaging plane (e.g. the sample may not be sitting perfectly flat in the sample holder). By calculating the two point correlations in space and time on the height-fluctuation field, we extract second-timescale changes in height with strong correlations with 100 $\mu$m in space. We note in SI movie and in fig. 1 G-H that these spatio-temporal dynamics are punctuated by rapid changes in height on short length-scales with single parts of the tissue dipping down on timescales which are intermediate between the single cilia beat frequency ($\sim$ 0.2 seconds) and the locomotive dynamics of the organism ($\sim$ 10's seconds). 

\section*{Model by construction of ciliary walking}

To digest our experimental observations, we adopt a model by construction framework which builds the complexity of underlying model in a step-by-step fashion constrained strongly by our experimental observations. The goal of this model by construction is to capture how the observed height dynamics arise from and feedback upon ciliary walking.

\subsection*{Ciliary shape dynamics are resilient to forcing}

A cilium under external load can be understood as the local state-space flow on a low dimensional manifold embedded within a much higher dimensional configuration space\cite{ma2014active}. This means that external stimuli will have the effect of pushing the cilium away from this manifold and the observed response can be understood as the sum of the relaxation back to the manifold mixed with the locally evaluated dynamics. In the presence of larger external load/force, stiffer directions away from this manifold will begin to play a larger role. If the manifold is very stiff against the environmental stimulation (e.g. stimuli will be unable to pull the state off the manifold, only along it) then one can parameterize the dynamics as single coordinate which keeps track of where on the manifold you are. We call this the phase and denote it as $\psi(t)$. 

The next leading order contribution which has been used to great success in capturing ciliary dynamics \cite{vilfan2006hydrodynamic, ma2014active, klindt2016load} is how far from the manifold the observed configuration is in a softer direction orthogonal to the phase, which we call the amplitude, $\rho (t)$.  

We extend this picture to propose a useful parameterization for the dynamics of a cilium interacting with a substrate. With this in mind, we propose a sufficiently simple parameterization to make progress while still illustrating the concept of replacing constraints in the softest direction away from the manifold with stiff degrees of freedom -- the phase-amplitude-curvature oscillator. Here we have a dynamical system for the time evolution of the amplitude, the phase and the static curvature of the cilium\cite{geyer_independent_2016}. We select curvature for two reasons, the first is curvature's critical role in interaction height as seen in experimental observations and the second is our understanding of ultrastructural stiffness \cite{hill2010force, gilpin_multiscale_2020}. 

The dynamics of this reduced dimensional system then takes the form of:

\[
\dot \rho = -k_\rho(\rho - \rho_o)  - \nabla_\rho \mathcal{E}
\]
$$
\dot \psi = \Omega - \nabla_\psi \mathcal{E}
$$
\[
\dot \Psi_o = -k_\Psi(\Psi_o - \bar{\Psi_o})  - \nabla_{\Psi_o} \mathcal{E}
\]

where $\rho$ is the amplitude, $\psi$ is the phase angle of the oscillator and $\Psi_o$ is the curvature of the oscillator as the leading order contribution to the bending energy which is written down as $E(\rho, \psi, \Psi_o | E_a)$ which we detail in the coming section.

\subsection*{Ciliary slipping-walking-stalling}

A cilium has been shown to exhibit an array of interactions with a substrate. Paramecium reverse their ciliary beatform in contact with a substrate \cite{wan_origins_2021, jekely_chemical_2021}, ciliary mechanosensing can enrich single-celled organisms near surfaces\cite{ohmura_simple_2018}, and algeal cilia have photoswitchable adhesion to surfaces \cite{kreis_adhesion_2018}. Here we focus on cilia which strongly interact with the substrate via a form of adhesion - whether hydrodynamic [see SI], electrostatic, hydrophobic or biochemical moiety. 

With much still to learn about how precisely placozoan cilia adhere to surfaces, we propose a simple formulation of adhesion energy constructed from the same principles underlying fracture mechanics and Griffith's criterion: bonding and debonding via an energy balance criteria. Using this simple principle, we can mix the geometry of the ciliary beat with substrate interaction and create a multistate dynamical system with a elastic-energy-balance defined switching criteria: on-the-surface and off-the-surface. We can write out this criterion by defining the 'energy' penalty of the ciliary beat being pulled away from the limit cycle and a parameter, $E_a$, which controls the adhesion energy between the cilium-point-of-contact and the surface. 

Using z-x plane coordinate system, we can write down the configuration energy of this system as:
$$
 \mathcal{E} (h, \psi) = \begin{cases} 
      E_{a}& \text{ if: } cos(\psi) > \frac{h}{\sqrt{\frac{2 E_a}{k}} + \ell_o}  \\
      \frac{k(h sec(\psi) - \ell_o)^2}{2} & \text{ else }
   \end{cases}
$$

We illustrate simple ciliary-tip trajectories in figure \ref{fig:fig2}c, by showing how phase angle stalling when in contact with the surface combined with tissue translation can give rise to cusp-like orbits punctuated by the transitions of entering and leaving the surface. 

We can compose these dynamics into a set of stochastic differential equations governing the evolution of the phase angle, $\psi(t)$ and the preferred height of the cell body, $h(t)$. These become:
$$
\gamma_h\frac{\partial}{\partial t} h = -ksec(\psi)(hsec(\psi)-\ell_o)*\delta_{Contact}
$$
$$
\gamma_\psi \frac{\partial}{\partial t} \psi = \Omega -kh^2sec(\psi)tan(\psi)(sec(\psi)- \ell_o/h) *\delta_{Contact}
$$
where the variable $\delta_{Contact}$ keeps track of whether the tip is in contact with the substrate or not. Recall that this state is controlled by the geometric criterion: $cos(\psi) > \frac{h}{\sqrt{\frac{2 E_a}{k}} + \ell_o}$. 

Equipped with a sufficiently simple ciliary beat, we start to classify the possible phases of ciliary-substrate interaction at the level of a single cilium. We begin to explore the parameter space of this model by studying a grid search across two input parameters: the ciliary tip force, $\Omega$ (which is related to the native frequency of the cilium), and the energy of adhesion, $E_a$. To characterize this collection of time-series using an array of summary statistics which correspond to experimentally observable outcomes: (i) the mean of the cell height, (2) the standard deviation of the cell height, and (3) the mean normalized period of the ciliary beat.  

In figure \ref{fig:fig2}D, we show these summary statistics for a given $\Omega$ for varying values of $E_a$. The combination of these three summary statistics reveal a broad crossover between pushing off the surface ('slipping') and being stuck to it ('stalling') which is in agreement with ciliary walking. 

The height dynamics for different parameters tell part of the story. Small ciliary tip forces quickly stall (as seen by the duty cycle $\sim$ 1) and pull the cell down toward $h = 0$. As the ciliary tip forces increase to an intermediate strength, the cells begin jumping around in height characterized by a high standard deviation between $0.2 < \Omega < 4$. Above this intermediate region, the cells push up to a high mean height with a relatively small duty cycle. 

Looking to the importance of the adhesion energy in figure, we identify an intermediate regime between stalled and low duty cycle which exhibits wild fluctuations in height. This broad crossover regime between stalled and slipping shows rich dynamics which demonstrate the opportunities for exploiting ciliary adhesion energy to not only colonize, but locomote upon life-rich ocean surfaces.  

\subsection*{The onset of ciliary stalling}

The complementary sharp transition from green to blue regions can be understood as a onset of the stalling discussed in figure 2E, but we can take this discussion one step further by studying a limiting behavior. This divergence in mean period can be solved for analytically in the limit where the point of contact does not slip along the substrate and the cell is set to move at a fixed speed, $v_{tissue}$, and height, $\tilde h = h_{tissue}/L_{cilia}$, propelled along by the rest of the tissue. In this limit, the problem becomes a geometric calculation where a circular trajectory interacts adhesively with a flat substrate as  seen in figure 2 B-C. In this limit, we find that the characteristic frequency of ciliary beating is:
\[
f_{cilium} = T_{cilum}^{-1} =  \frac{1}{2}\left[\underbrace{\frac{\pi - asin(\tilde h)}{\Omega}}_{\text{time off-surface}} + \underbrace{\frac{\sqrt{1-\tilde h^2}}{v_{tissue}}}_{\text{time on-surface}}\right]^{-1}
\]

As the preferred tissue speed diverges, $v_{tissue}\rightarrow \infty$, the frequency of ciliary beating approaches the natural frequency of the cilium, $f_{cilium} \rightarrow \Omega/2\pi$. As the normalized height approaches one, the frequency also converges on the natural frequency of the cilia. As $v_{tissue}$ drops in the limit of no slipping, the cilium period diverges as all the time is spent stalled in contact with the substrate (e.g. no advance in ciliary phase, $\dot\psi \rightarrow 0$). 

This limiting case reveals the intricate interplay between height, activity and local tissue speed established by the population of cells around the single cilium under study. With it, we have shown how the effective ciliary beat frequency can span orders of magnitude and suggests that ciliary walking in this limit can be understood as a dynamic interplay of these three variables. We posit that this interplay underlies some of the richness of behavior in placozoa which we continue to explore in the coming sections.

\section*{Walking cilia are sensitive to external stimuli}

Cilia combine sensing and actuation into a single organelle bypassing the need for long-range information transmission and storage \cite{gilpin_multiscale_2020}. A walking cilium can be subject to a large number of stimuli which span the spectrum biochemical modifications to the internal state to external force stimuli subjected to the cell body. Here, we study how a model cilium responds to external stimulus applied to the cell body in the form of a force in the tissue plane and a force out of the tissue plane. 

In the limit that the force is small compared to the rest of the forcing, we can study the response of the system by monitoring it change in phase, $\Delta \psi$ in response to the stimulus, $I$. This small forcing limit corresponds to one of the simplest forms of susceptibility which varies with internal state and goes by the form the phase response curve $Z(\psi)$. Work in the neuroscience of oscillating neurons has shown that the inference of the phase response curve can be measured through looking at integral of the spike triggered average. Here, we are simply able to use the geometry of the situation to argue that the dominant contributions for each force looks like a truncated cosine (out of plane) and sine (in plane) respectively. 

We begin by plotting the solution to the phase angle evolution where the angle subtended by the surface is reflected by a light blue region. In the limit where the ciliary driving is well above the stalling transition (but is still in the 'walking' crossover phase) the cilium takes the form of a surface modified oscillator. However, when we study the stalled regime, the cilium advances until it is stuck trying to exit the surface but is held in place by the adhesion energy. 

We take both of these systems, the surface modified oscillations and the stalled oscillator and apply a small of in-plane-force stimulus for a short pulse. The walking cilium's response is strongly dependent upon when the stimulus arrives and on average is perturbed a small amount (see figure 3C)

The stalled cilium on the other hand exhibits a sensitive response reminiscent of an excitable system. The stalled cilium sits in a position where the ciliary tip force is canceled by the substrate adhesion and both forms of stimulus (in plane and out of plane) have a positive phase response. This means that either form of stimulation can help the ciliary tip to overcome the trapping potential and to sweep out a ciliary 'step' where the tip peels of the surface, completes a recovery stroke and then attaches back to the surface at the surface-entry phase angle. For this reason, we use the term 'step excitability'; to describe the role of external forces in kicking the stalled cilia off the surface (see figure 3D-E).

By studying the response as a function of different size external stimuli, we reveal a threshold-like function above which the cilia is able to take a step and below which it remains on the surface and relaxes back toward the stalled position with an exponential relaxation. This response pattern is reminiscent to other nonlinear dynamical systems which are far from equilibrium such as the FitzHugh-Nagumo equation \cite{fitzhugh_impulses_1961} except that this dynamical response is chemomechanical more similar in flavor to mechano-chemical excitability observed in the actomyosin cortex\cite{boocock_theory_2021, armon_epithelial_2020}. 

To understand the impact of the stimulus in dynamical systems, the simplest metric would look something like the distance between the state before the stimulus and the state after the stimulus: $\psi(t_{\star} + T) - \psi(t_{\star})$. However, this suffers from a strong sensitivity on the local speed even without the stimulus (e.g. the oscillating case never stalls). We instead propose a metric which is not as sensitive to speed of the state evolution after the stimulus as would be above. As a measure for stimulus response, we study the state difference at the same timepoint between two parallel systems, one which is stimulated at time $t_{\star}$ and one which is not. This reveals how much the actual trajectory was impacted by the presence of the stimulus 
\[
\chi(I) = \psi(t_{\star} + T|I\neq 0) - \psi(t_{\star} + T |I=0).
\]

A grid search for a fixed amplitude and time of the stimulus shows that the maximum response is found on the boundary between stalling and walking ciliary phases (see figure 3E). This observation is consistent with the idea that the susceptibility is maximized at transitions between qualitatively distinct dynamics. This result suggests that placozoa might have some incentive to tune their ciliary tip driving force to closely match the energy of adhesion in exchange for tuning the systems locomotive sensitivity to external stimuli. 

\subsection*{Experimental signatures of step excitability}

In the search for step excitability, we looked at the stalled tissue, which has been reported previously during feeding\cite{smith2015coordinated}. By imaging the single cilium resolved cross-section and taking a digital bandpass filter to improve the ciliary cross-section contrast, we can observe signature of ciliary steps both in local tissue motion and in actual ciliary movement. To highlight these dynamics, we can take a registered image difference, $I(x,y,t+1) - I(x,y,t)$, the amplitude of which we call image disturbance to highlight the time dynamics of motion locally.

In figure 3F we show a $50 \mu m$ section of tissue in the stalled state. In the initial frame, a single cilia takes a step causing the tissue above to move. A full second later, the a region nearby shows a large number of cilia ($\>5$) taking a synchronized step before settling back down to a reasonably quite state. Another second later, a single cilium takes a step starting the process over. The time dynamics of this image motion over this small section of tissue illustrates two qualitatively important phenomena: (1) a stepping cilia can generate force sufficient to cause an 'avalanche' of steps in response and (2) the dynamics are excitable but do not grow to system size during the stalled state.

\section*{Collective phase dynamics and synchronization of ciliary walkers}
Here-to-fore, we have demonstrated that a single cilium (with dynamics governed only by an emergent manifold sweeping out in shape space) exhibits a rich dynamical space interacting with a substrate. Yet, one might expect that the organism's locomotion may be captured by something more than the simple sum of its cilia. Should we expect walking cilia to synchronize? Or can the phase dynamics be successfully neglected? 

To explore the role of synchronization in the dynamics we begin by looking deeply at the experimental data available before reconciling these observations limiting solutions to the phase difference equations (1) the limit of weak surface influence and (2) the limit of strong surface influence. We report the discovery of a small region of stable synchronization which disappears as the height of the tissue exceeds the length of the cilia.

\subsection*{Experimental observations support a height mediated synchronization transition}

Returning to our tracking microscope focused $~5 \mu$m below the tissue plane, we can directly visualize the dynamics of ciliary steps at high spatio-temporal resolution by tracking the ciliary cross-sections which show up as small dark occlusions. Across the full field of view, we can extract regions which are higher relative to the mean and lower relative to the mean and study the dynamics of these occlusions using hand-stitched segments of short ciliary cross-section trajectories. 

The experimental measurements of step synchrony are height dependent. At low heights, the ciliary steps have strong spatially local synchrony, but as the tissue rises, the very same set of cilia lose step synchrony over a few second period. 

Zooming out and observing the dynamics across larger ensembles of cilia, we can measure a discrete analog of the phase angle density $\rho(\psi,t)$ which is the local density of cilia at a given phase angle and time $N_{step}(t) = \int_{\psi_{on}}^{\psi_{off}} d\psi' \rho(\psi',t)$. We can gain direct access to this function by asking about the percentage of stepping cilia at every point in time. If the cilia are tightly clustered together in phase angle space, we would expect to see a large number of ciliary steps simultaneously which would manifest in a large periodic fluctuation in the number of stepping cilia over time. Perfect synchrony would look like a box function where all the cilia begin stepping at the same moment then enter the surface (sending $N_{step} = 0$) at the same moment. Therefore more synchrony manifests itself as more variation in the number of stepping cilia at any one moment. By showing that at high height, the number of stepping cilia fluctuates only minimally around $8\%$ stepping at any one period of time, this lends support that higher heights exhibit less synchrony. 

In brief, we see strong experimental evidence that the transition to  synchronization is height-dependent. The theoretical model in the next section was developed to further study the physical origins (and limits) of this phenomenon.

\subsection*{Synchronization of two walking oscillators}

%[Computation non-biology: excitability of stepping oscillators. Direct mapping of presence of synchronization to phenomena -- metachronal, height controllability (hidden fields)]

A natural place to begin the study of the collective phase dynamics of walking cilia is by studying the pair of walking cilia. 

We begin by deriving the entrainment equations for how a ciliary oscillator couples to an existing external signal of the form of a coupled trigonometric functions. In the small signal and large amplitude stiffness approximation, let's write the single oscillator phase which is subject to two forms of external stimulus:
\[
\partial_t\psi = \Omega + Z_1(\psi) I_1(t) + Z_2(\psi) I_2(t) 
\]

In the context of a ciliary walking oscillator we interpret the first form of stimulus $I_1(t)$ as the time varying force in the tissue plane. $F_x$. The second stimulus then becomes the time varying force in the vertical plane (out-of-plane force in the $h$ direction), $F_h$. These two orthogonal channels of information can arise from either a common source or different sources. Here, we apply this line of thinking to study the response of a single oscillator in response to a common signal coming from another walking cilium oscillator.

As we showed previously, we can approximate the qualitative shapes of these phase response curves as:
\[
Z_1(\psi) \sim sin(\psi)
\]
\[
Z_2(\psi) \sim cos(\psi)
\]

We consider the case where the signal is coming from an identical type of phase oscillator and is coupled through the vertical and in-plane forces the oscillator generates. Note that this generator oscillator is not yet effected by the oscillator under study and thus this falls into a study of the oscillator entraining to an external signal. In a previous section, we showed that the forcing arises from a walking oscillator takes the form of:
\[
F_x(\psi) = \omega cos(\psi) + \nabla_{\rho} E sin(\psi)
\]
\[
F_h(\psi) = -\omega sin(\psi) - \nabla_{\rho} E cos(\psi)
\]
where $\nabla_{\rho} E$ is the gradient of the free energy with respect to the amplitude degree of freedom.

Since the phase angle of this signal generating oscillator is fixed in time, let's map $\psi = \omega t$ and apply this to the original phase evolution of the coupled oscillator. After applying the classic angle-difference relations, we find:
\[
\partial_t \psi = \Omega - \omega cos(\psi-\omega t) + \nabla_{\rho} E sin(\psi-\omega t)
\]

We recall that $\nabla_{\rho} E$ is comprised to two coupled contributions: the limit cycle stiffness and the energy of surface adhesion. These two contributions define the relative magnitude of the force in the phase angle coming from the phase angle compression effect arising from $E_a$ and the stretching effect arising from the force bump at $\psi=0$ and controlled by $k$. When the phase density compressing effect dominates, high $E_a$, the sign of $\langle\nabla_{\rho} E\rangle_t$ will be negative. When the phase stretching effect dominates at high $k$ and low height, the $\langle\nabla_{\rho} E\rangle_t$ will be positive. The positive sign will correspond to synchronization and the negative sign will correspond to de-synchronization regime. 

\[
\epsilon(k, E_a, h)
\]

A simple way to understand $\epsilon$ is how much on average, the two oscillators are pushed together or pulled apart over a cycle. We solve for this emergent feature numerically and find that the average convergence crosses over from negative to positive at a height which is controlled by the microscopic parameters. Since the force arising from the surface interaction is an odd function around vertical, $\psi = 0$, the two halves can be summarized as a time-averaged value [see SI for time average from geometric average] of how much the cilia is preferentially pushed toward or away from $\psi = 0$. The two contributions coming from the adhesion and limit cycle stiffness play opposing roles, with the adhesion pulling the cilia back together and the limit cycle stiffness pushing them apart. The crossover in the dominance of these two terms is reflected by the change in sign of the average convergence $\epsilon$.

We can then relax the assumption that the driving oscillator is not altered by the state of oscillator under study, but opening our definition of the system up to two oscillators and zero external input. The parity of this coupling allows us to write the full dynamics of these oscillators as:
\[
\partial_t \psi_i = \Omega + \delta_s (\psi_j) \left[-\Omega cos (\psi_i - \psi_j) + \epsilon sin(\psi_i - \psi_j)\right] 
\]
where $i\neq j$ and $i,j \in \{1,2\}$ and the surface contact nonlinearity is tracked by the binary variable $\delta_s(\psi)$.

To build intuition for these dynamics, we first neglect the surface non-linearity and transform into a rotating reference frame where we can study the fixed points of the phase angle difference by subtracting $\partial_t \psi_i - \partial_t \psi_j$. It is worth-while to note that the loss of the surface nonlinearity is an approximation designed to tie this analysis to existing techniques and only applies in the limit where the entire limit cycles remain in contact with the surface.

\[
\partial_t \Delta = \delta\Omega + -\delta\Omega cos (\Delta) + 2\epsilon sin(\Delta)
\]
where we define: $\delta \Omega \equiv \Omega_i-\Omega_j$.

We can plot this one dimensional dynamical system for various values of $\epsilon/\delta\Omega$ to show the crossover from a stable fixed point at $\Delta=0$ to an unstable fixed point. 

The resulting curve in fig 4k reveals two simple lessons: (1) for $\epsilon$ the phase angle difference $\Delta = 0$ is stable will result in synchronization mediated through the phase response curves to stimulus generated from a common source and (2) when $ \epsilon <0$ the $\Delta=0$ fixed point becomes unstable resulting in a preferred phase lag between two coupled oscillators. We should be careful interpreting much about the preferred phase lag until we show the role of the contact nonlinearity switching in the next section. 

\subsection*{Incorporating the surface nonlinearity}

Relaxing the assumption that allowed us to neglect the surface interaction nonlinearity requires a new technique to study the fixed points of the recursion map (Poincur\'e section). We cannot transform into a rotating frame of reference because the symmetry is broken by the presence of the surface. This map can be derived in the limit of $\delta \Omega << \Omega_i$ by keeping track of all the possible states the two oscillators can be in and integrating the dynamical system over those trajectories [see SI section 3] until we return to the start.
\[
\Delta_{t+1} -\Delta_t = \sum_{i \in \{states\}} \partial_t \Delta_i \tau_i
\]
This map is analogous to watching a closed circuit race (e.g. Formula One or Nascar) and taking a snapshot of the relative position between car number one and two each time car number one crosses the start line.

Denoting the possible states with pairs of zeros and ones allows us to keep track of the two leading order projections (in the limit that the difference is angular speed is small relative to the angular speed itself):
\[
... \rightarrow 00 \rightarrow 01 \rightarrow 00 \rightarrow 10 \rightarrow ...
\]
\[
... \rightarrow 00 \rightarrow 01 \rightarrow 11 \rightarrow 10 \rightarrow ...
\]

With simplification our return map takes the form of:
\begin{widetext}
\[
\mathcal{M}(\Delta) \approx \left(\frac{2\pi -  \Delta\psi}{\Omega_2} + A(\Delta)\right)   \delta \Omega  
- A(\Delta) \delta \Omega cos \Delta + 
 A(\Delta)\epsilon sin \Delta
\]
\end{widetext}

Where we define the quantity $A(\Delta) \equiv \left( \frac{\Delta \psi - \frac{\omega_1}{\Omega}\Delta}{\omega_\star} \right)$ and becomes zero when $A(\Delta) <0$. This quantity measures the amount of time spent in the "11" state with both cilia in contact with the surface.

The primary modification due to the time off surface has the effect of shifting the fixed phase angle difference away from zero. In the limit that $\epsilon \Omega_2 >> 2\pi-\Delta\psi$, the fixed point returns toward $\Delta = 0$ with small values of $\epsilon$ having no solution.

We can approximate the numerically discovered map between $\epsilon$ and $h$, height, as:
\[
\epsilon(h) \approx (h-h_{cross})(h_{max}-h)(h<h_{max})
\]
where $h_{cross}$ is the height at which the value of epsilon crosses zero (functions of the micro-parameters of the problem) and $h_{max}$ is the height at which the cilia no longer contacts the surface defined as $\sqrt{\frac{2E_a}{k} + \ell_o}$. This approximation holds nicely for a wide range of microparameters, but can be replaced with an explicit numerical integral to achieve qualitatively consistent results. 

By injecting both the numerical integral and the simple approximation (which aids in description), the non-dimensional parameter can be replaced with the physical parameter corresponding to height and the stability of the recursion map for the phase difference between two oscillators can be viewed as a function of the observed tissue height in figure 4M. 

Critically figure 4M shows that there is a window in height from the surface (at intermediate heights) in which ciliary in-phase synchronization is preferred in the absence of noise. It is also noteworthy that at very low heights the stability of the $\Delta\rightarrow 0$ fixed point changes and the stable phase separation reaches a finite value which approaches antiphase synchronization in the limit of weak surface interaction. 

To build intuition for the increased importance of the difference in the preferred native frequency between the two oscillators $\delta\Omega = \Omega_i - \Omega_j$ (which arises from the integral over the full cycle), we can plot up the null-clines of our equation for increasing values of $\delta\Omega$. This reveals in figure 4N the transition from stable in-phase synchronization to an unstable phase difference similar in flavor to the phase slips of the Adler equation \cite{gilpin_multiscale_2020}. It is noteworthy that the low height anti-phase synchronization is more robust against increasing $\delta\Omega$ than the intermediate height in-phase synchronization. 

\subsection*{Synchronization is preserved at intermediate height in all-to-all coupling model}

We study the full dynamics of the ciliary walking model numerically by simulating $N=500$ phase-amplitude-curvature oscillators and observing the synchronization of long-time steady-state. We complete $2e3$ numerical experiments across two parameter space planes: (1) ciliary driving, $\Omega$ versus energy of adhesion, $E_a$ and (2) limit cycle stiffness $k_{LC}$ versus the energy of adhesion $E_a$. We find that ciliary synchronization is expected close to the boundary between walking and stalling and in the regime where $\epsilon > 0$. 

To develop a simple take-away, we summarize this data by comparing the synchronization against the self-organized height and find that in bulk across $2000$ sets of parameters, that the synchronization occurs at an intermediate height which is primarily tuned by the energy of adhesion. This strong correlation suggests that height may be a useful proxy for understanding when ensembles of walking cilia may synchronize. This result can be understood in the language of the weak surface influence model where $\epsilon$ crosses over from negative to positive at an intermediate height. 

\section*{Phase averaging reveals a multistable height-activity relationship}

In the high height regime where synchronization is strongly disincentivized, it become advantageous to average over the phase degree of freedom to study an emergent height versus active forcing relationship. This approach can be justified on two grounds: (1) separation of timescales gives rise to multiperiod dynamics and (2) ensembles of desynchronized cilia locally will take the time average of the phase. A careful application of this technique will derive an emergent relationship between height and active forcing.

\subsection*{Cartoon/analogy to pitch/tires/wheelies}
The high static curvature component \cite{geyer_independent_2016, klindt2016load} of the placozoan cilium beat is essential to understand the observed dynamics ($\psi_o \sim 0.02$ radians/$ \mu m$). Critically, this high static curvature enforces an offset (in the $\hat x$ direction) between the position where the cilium touches the surface and the position where the cilium attaches to the cell at the basal body. The mechanics of torque balance in this situation can be well understood via analogy with a motorcycle 'wheelie' in which the center of mass is lifted when the driven tire torque exceeds the torque on the center of mass due to gravity and the offset from the back tire. In this way, high application of force on the surface results in a torque which lifts the tissue up. Whereas, if the force on the tissue (in the direction of motion) exceeds the torque generated by the ciliary interaction with the substrate, the tissue will be pushed down to a lower height.

Through this mechanism the static curvature is essential for understanding the height dynamics of a ciliary walker. We can derive this relationship by recognizing that a force in-balance between the forces acting in the tissue plane and the locomotive forcing will cause a torque rotating the cell around the point of surface contact. For a fixed length, this means that the change in height takes for the form of:
\[
\partial_t h = - sin(\Theta) \sum_i \ell_i \times F_i
\]
where we have defined $\Theta$ as the angle between the point of contact and the cell body. 

It is important to remember that changes in height also have the effect of changing the relative distance between the point of contact and the tissue. With a constant static curvature, the mean ciliary shape sweeps out a portion of a circle. This means that with reduced height the angle between the point of contact and the cell body will decrease $sin(\Theta(h)) \sim \sqrt{r_o^2 - h^2}$, where we define $r_o$ as the radius of the fictional circle swept out by the static curvature $r_o \sim \psi_i^{-1}$. At lower heights, the average point of contact is pushed closer to the cell body which has the effect of reducing the contributions of this term. As the height reaches the radius of the fictional circle, $r_o$, the curve flattens to a slope of zero.

We write this as:
\[
\partial_t h = L(h) \sum_i F_i + F_h
\]
where $L(h) \equiv \ell \sqrt{\psi_o^{-2} - h^2}$ and $F_i$ are the forces in the tissue plane applied to the cell-body and $F_h$ are the forces applied to the cell-body in the height ($\hat z$) direction. 

This torque mediated coupling between the in-plane forces and the height leads to a feedback between locomotive forcing and height control. 

\subsection*{A walking cilium generates a locomotive force maximum at an intermediate height}
We can employ this separation of timescales combined with the limit of zero-synchronization to begin studying the cycle-averaged locomotive force created by a walking cilia. This approximation removes the overhead of keeping direct track of the phase angle of the ciliary oscillator and allows us to study the emergent dynamics on the multiple cycle timescales similar to that observed in the 100 $\mu$m length and second time-scale dynamics observed in the organismal locomotion (see: figure 1D).

The force acting on the cell at any one point in time is determined by the configuration:
$$
F_{cilia, x}(h, \psi) = \Omega sin( \psi + \pi/2) + k_c(\ell(\psi) - \ell_o)\hat r(\psi')\cdot \hat x 
$$

Writing this in trigonometric functions applied to the phase angle, this becomes:
$$
F_{cilia, x}(h, \psi) = \Omega cos \psi + k_c(\frac{h}{cos\psi} - \ell_o)sin\psi
$$

The time average force is then equal to the path integral of this force as swept out by the configuration evolution $h(t)$ and $\psi(t)$.

The dynamics which generate $h(t)$ and $\psi(t)$ are:
$$
\gamma_h\frac{\partial}{\partial t} h = -\nabla_{h}E
$$
$$
\gamma_\psi \frac{\partial}{\partial t} \psi = \Omega -\nabla_{\psi} E
$$

If we assume that $\nabla_\psi E << \Omega$ (i.e. the internal dynamics of the oscillator governs the beat more-so than the weak environmental interaction), the trajectory is well approximated as a constantly growing $\psi$ with time.

This approximation replaces our time average as a phase average.
\begin{widetext}
$$
\langle F_{cilia, x}(h) \rangle_{T} \approx \langle F_{cilia, x}(h) \rangle_{\psi} = \frac{1}{2\pi} \int^{\psi_{snap}}_{-\psi_{snap}} d \psi' \left[ \Omega cos \psi' + k_c(\frac{h}{cos\psi'} - \ell_o)sin\psi'\right]
$$
\end{widetext}

Where $\psi_{snap}$ is defined by the energy balance between elastic and adhesive energies  $E_e \sim \frac{k_c\Delta \ell^2}{2}$ and $E_a$. 
$$
\psi_{snap} = arccos\left(\frac{h}{\sqrt{\frac{E_a}{k_c}} + \ell_o} \right)
$$

Multiplying through, this integral accepts an analytical solution:
$$
\langle F_{cilia, x}(h) \rangle_{\psi} = \frac{1}{2\pi} \int^{\psi_{snap}}_{-\psi_{snap}} d \psi' \left[ (\Omega - k_c\ell_o) sin \psi' + k_ch tan\psi' \right]
$$

Recall that $\int dx tanx = -\log|cos| + C$ and that $\int dx sinx = cosx + C$. This then becomes:

$$
\langle F_{cilia, x}(h) \rangle_{\psi} = \frac{1}{2\pi} \left| \left[ (\Omega - k_c\ell_o) cos\psi - k_c h \log|cos \psi| \right] \right|_{-\psi_{snap}}^{\psi_{snap}}
$$

Evaluating this at our snap on values gives:
\begin{widetext}
$$
\langle F_{cilia, x}(h) \rangle_{\psi} = \frac{1}{\pi} \left( (\Omega - k_c\ell_o) \frac{h}{\sqrt{\frac{E_a}{k_c}} + \ell_o} - k_ch \log\left| \frac{h}{\sqrt{\frac{E_a}{k_c}} + \ell_o}\right| \right)
$$
\end{widetext}

We can select reasonable relative values to plug into this equation and can plot up the average force as a function of the cell height.

The observation (see figure 5B) that the general shape of this curve is well approximated by $f(h) \approx f_{max}(1 - \beta_{sen}(h-h_{max})^2$ allows us to reduce the number of parameters we use to study these dynamics. We can translate the parameters of adhesion energies, ciliary forces and stiffness into a more phenomenologically universal formulation characterized by a leading order Taylor expansion. 

\subsection*{Emergent height regulated activity}

The essential features of the coupling between height and force is that the curve is non-monotonic and concave down with a maximum value at a finite height, $h_{max}$ from the surface. Motivated by this generic property of $v(h)$, we approximate this relationship as a second order Taylor expansion around $h_{max}$. The calculation then begins by finding $h_{max}$ using knowledge that it is an extremum of the equation above. We can take this derivative with respect to h to find:
\begin{widetext}
\[
\partial_h \langle F_{cilia, x}(h) \rangle_{\psi} = \frac{1}{\pi} \left( \frac{(\Omega - k_c\ell_o)}{\sqrt{\frac{E_a}{k_c}} + \ell_o} - \left(k_c \log\left| \frac{h}{\sqrt{\frac{E_a}{k_c}} + \ell_o}\right| + k_c \sqrt{\frac{E_a}{k_c}} + k_c\ell_o \right)\right)
\]
\end{widetext}

We can approximate it with the second order expansion around the max:
$$
 v(h) \approx v_{max} - \beta_{sen}(h-h_{max})^2
$$

One of the essential features of the observed self-organized height regulation is the interplay of the cycle averaged force and the torque between the point of substrate attachment and the cell body. When the cell body is offset from the point of adhesion, this system analogously mimics the back half of a motor-bike. When the rider hits the accelerator applying power to the back tire, the bike can lift up into a 'wheelie'. An analogous phenomena occurs when the ciliary force is stronger than the in-plane, elastic force acting on the cell. These forces apply a torque to the extended body between the cilia and the cell causing it to rotate in the X-Z plane (z, corresponding to height and x corresponding to the direction of the ciliary forcing). Through this mechanism, the system arrives at a self regulating height control which can be modified by either a change to the force applied to the cell or the force the cilia is exerting on the substrate. 

The two new terms in this equation are the gradient of the elastic free energy (which can take a complex form of an active rod-bending under external constraint\cite{gadelha2010nonlinear}) and the external force applied to the cell (could be applied either by environment or elastic coupling to neighboring cells).

Following the pitch equation from torque and force balance $\ell[h] \sim h$ arises from the assumption that our phase-amplitude-curvature oscillator does not buckle under sufficiently small changes in height. This compounds into the following equation governing the height evolution:
$$
\gamma_h \frac{\partial}{\partial t}h = L[h]\left( (v_{max} - \beta(h-h_{max})^2)\gamma_o - F_{ext}\right) - k_h (h - h_o)
$$

This equation represents our effort at reducing the monumental complexity of a cilia interacting with a substrate into a dynamical system which agrees qualitatively with the dynamics we observe in both experiment and simplified numerical work. The goal of the next section is to study this emergent simplicity  for interesting dynamics akin to the way one would study a nonlinear differential equation.

\subsection*{A pitch instability generates excitable dynamics}

The two way coupling between height and locomotive forcing gives rise to a pitch instability for sufficiently high sensitivity to change in height $\beta$. The dynamics of tissue height takes the form of:
$$
\gamma_h \frac{\partial}{\partial t} h = L(h) \left(v(h) -\nabla_x E_{tissue} - F_{external}  \right) - \nabla_h E_{elastic}
$$
With 
$$v(h) = \begin{cases} v_{max}-\beta(h-h_{max})^2 & \text{while on surface} \\
0 & \text{while off surface} \end{cases}$$

We can employ a simple method to find the stable points for each height. We initialize many uniformly sampled initial conditions for each external force magnitude $F$ and allow the dynamical system to converge toward the stable fixed points in the limit of many time-steps. To get the unstable fixed points we complete the same procedure for the negative of the differential equation. This gives us a simple diagram of the fixed points in height and injection of ciliary force as a function of the external force pulling on the cell.

\subsubsection*{Dominated by external force we discover an emergent bi-stability}

In figure 5E, we begin by neglecting the force contributions from the tissue, analogous to considering the dynamics of a single cell to external stimulus in the form of a pull (force clamp) in the plane of the tissue. We study these response dynamics as a function of the magnitude and direction of the pull force. 

The curves in figure 5F show that for a choice of the height velocity relationship, our governing equations exhibit an emergent bistability at intermediate external forcing. This predicts that if an external force is applied to a cilia in the direction of its motion that a new low-height, low-speed state will emerge and become stable. At low force, only the high-height state is stable and at high force only the low-height state is stable. This suggests that the application of a large force in the direction of motion of the ciliary walker will cause it plunge down toward the substrate. Above a critical $\beta$, the dynamical system undergoes a supercritical bifurcation. Above the critical value of $\beta$ one stable fixed point turns into two stable fixed points separated by an unstable fixed point. 

\subsubsection*{Dominated by tissue forces, the high state can begin feedback oscillation}
There are two injections of planar force coming into a walking cell which we can delineate into (i) external stimuli and (ii) coupling forces to other cells. In the section above, we focused on the regime in which the external stimuli dominates the input into the cell. In this section, we will consider a reduced order model in which the elastic cell-to-cell interactions is the dominate contribution to the dynamics. 

We parametrize the motion of the tissue by setting it to a preferred speed in units of $v_{max}$, $\nu$. By elastically coupling the cell to this translating tissue, we define the force arising from the tissue as an integrated disagreement of the form: $-\nabla_x E_{tissue} = k_{tissue} \int (v - \nu) dt$. When applied, this gives us a new governing equation for the height which takes the form of:
\[
\gamma_h \frac{\partial}{\partial t} h = \ell(h) \left(v(h) -k_{tissue} \int_{-\infty}^{t} (v(h(t')) - \nu) dt'  \right) - \nabla_h E_{elastic}
\]

This has a form reminiscent of the integral contribution of a classic PID controller decorated by a non-uniform (height dependent) proportionality, $\ell(h)k_{tissue}$, an offset (such that $v(h) -k_{tissue} \int_{-\infty}^{t} (v(h(t')) - \nu) dt' = 0$) and a simple stabilizing potential which penalizes large deviations from the height of the rest of the tissue, $h_o$. The integral term often has the effect of increasing oscillation and instability around the fixed point with the benefit of removing small but persistent systematic errors.

\subsection*{Feedback between the tissue locomotion and cellular locomotion controls switching dynamics}

Taking the lessons from the past section on the interplay between tissue elasticity and in-plane forces, we can study a two-D dynamical system in which heteroclinic orbits control the switching dynamics between the high and low state. 

This leading order approximation of the dynamics of the in-plane force for a fixed velocity of the tissue gives rise to a dynamical system which mimics the similar form to the Fitzhugh-Nagumo model where the recovery variable, $W$, is mimicked by the elastic force. The effective dynamics looks a little different, but takes the form of:
$$
\partial_t h = F(h) + h(-F_x - I) - F_h
$$
$$
\partial_t F_x = -k(v(h)-v_{ext}) - rF_x
$$
with 
$$
v(h) = v_m - \beta(h-h_m)^2
$$

In figure 5D, we can define a phase portrait for the this system of coupled nonlinear dynamical systems and show that the stability is preserved.

\subsection*{Is bistability preserved under other forms of closure?}

Thus far we have treated the dynamics of the cilium as a system under external forcing. The intuition, we gained from this work holds well in the limit of slow changes to the external fields relative to the dynamics of the cilium.

For completeness, we study the dynamics of this model under a new form of closure, where tissue velocity and height is not imposed by an external system but is instead closed through direct reference to the dynamics of an ensemble of these dynamical systems. The simplest fashion by which to accomplish this is an all-to-all coupling model in which $N$ can be written as a zero-spatial dimension model by placing them all in a bath with each talking to every other one. The tissue height and velocity are then updated as the mean of the collection of the $N$ dynamical systems. 

In figure 5H, the dynamics of 10 randomly selected 'cells' in the all-to-all model (with N=1000) are plotted showing both transients with a great deal of switching between high and low but also long time solutions which preserve the bistability as shown by $\sim 15\%$ of the cells diving down toward the surface with the rest finding a resting height high off the surface (but still in walking contact).

These results lend greater support the conclusion that the observed bistability is not an artifact of our choice of model closure but instead a more generic result.

\section*{Multistability and excitability in the animal}

We next look to the animal primed by our results which suggest that a ciliary walker at sufficiently high height from the surface (where synchronization may be successful neglected) can be seen as a multistable system with a stable high tissue height and low tissue height separated by an unstable separatrix. 

In a well-fed environment, \textit{T. Adhaerens} will periodically transition from walking to feeding by pressing itself up against the substrate and secreting digestive enzymes\cite{Ueda1999DynamicAdhaerence,smith_coherent_2019, fortunato_social_2019}. 

We directly observe this transition in, N=23, organisms at various degrees of magnification. At sufficiently spatio-temporal resolution (0.1 sec, 0.8 $\mu$m, across mm field widths), the transition is often punctuated not simply by a change in local opacity but also a change in height corroborated by the flow being ejected from underneath the organism. Visible in a dark field imaging modality, this height transition travels as a propagating wave at $100 \mu$m per second. This rapid mechanical snap to the surface is occurring in an animal without a central controller. 

In this section, we show that while walking mediated height switching is not the whole story, that in a particular regime of parameter space, the dynamics of a chain of phase-averaged ciliary walkers can support traveling switching waves. We consider various possible fashions in which this mechanical switching can complement known mechanisms in placozoa \cite{Senatore2017NeuropeptidergicSynapses.,romanova_sodium_2020} and other simple marine life\cite{wan_origins_2021}.

\subsection*{Up-to-down transition, feeding}

To understand the limit where tissue height transitions are fast, we need to relax the assumption that the tissue height is slowly evolving. This can be accomplished by writing the dynamics of the local tissue height using the tissue velocity integration $F_x = k_x\int dt \left[v(h) - v(h_o)\right]$ and the stiffness in the plane orthogonal to the substrate $F_y = -k_y (h- h_o)$. This gives rise to the three dimensional dynamical system of the form:
$$
\partial_t h = \ell h (v(h) - F_x) + F_y
$$
$$
\partial_t F_x = k_x \left[v(h) - v(h_o)\right] - r F_x
$$
where $r$ is the rate of stress relaxation in the viscoelastic tissue, and
$$
\partial_t h_o = -\tau(h_o-h)
$$

To capture the qualitative character of this system, we can employ a separation of timescales by assuming that the stress in the tissue plane $F_x$ relaxes to its steady state instantaneously on the timescales of the other dynamics. This assumption allows us to reduce this 3 variable dynamical system into a 2 dimensional dynamical system where the recovery variable takes the form of the tissue.

Solving for the steady state stress in the tissue plane in terms of height:
$$
F_x^{ss} = \frac{k_x}{r} \left[v(h) - v(h_o)\right]
$$

which can be plugged into the original system to find:
$$
\partial_t h = \ell h \left[ \left(1-\frac{k_x}{r}\right) v(h) - \frac{k_x}{r} v(h_o)\right] - k_y (h-h_o)
$$
$$
\partial_t h_o = -\tau(h_o-h)
$$

In a recurring theme, we find that for this 2d cross section of the dynamics, heteroclinic orbits in ciliary walking govern the fast transition between the two stable heights within the tissue. In this low dimensional model, we find two parameters which control how compliant the cell under study is attached to the rest of the tissue in the tissue plane $k_x$ and out the tissue plane $k_y$. 

Plotting up the null-clines as a function of $k_x$ shows (figure 5D) transition from a bistable system as low $k_x$ to a oscillating point at high height for large $k_x$. We interpret this transition as analogous to the oscillating instability in a PI loop when the integration term grows too large (see SI on height dynamics coupled to fixed tissue speed). 

Varying $k_y$ has a much more simple effect of compressing the null-clines down closer to the line $h_o = h$. This flattening has the effect of making the nearest point on the hetero-clinic orbits much closer to the rest configuration making knocking the system into switching trajectories more easily. However, there is a competing effect of $k_y$ which stabilizes the system in the high state by reducing the magnitude of the response to the same stimulus. The balance of the

\subsection*{Controlling the dynamics with other forms of stimulation}
We observe traveling height snap waves initiating rapid transitions between the high state and the low state (see figure 6A-B). Next, we ask: does the minimal bistable dynamics model of a phase averaged walking cilium support traveling waves of heteroclinic orbits initiated by stimulus?

We note that the observed traveling is consistent with the dynamics of the upper epithelium and that the minimal hypothesis is that these excitable wave dynamics on the top of the organism can couple into the bottom through mechanical or chemical coupling. There are two dominate fashions through which this coupling can occur: 1) external force acting on the tissue (in alignment with local ciliary orientation) and 2) through chemical signaling which modify v(h) (such as calcium or neuropeptide mediated ciliary reversal). 

To study the effect of both of these forms of stimulus, we developed a 1D chain model (with periodic boundary conditions) in which we excite the center cell via a stimulus for a given start time and duration (figure 6F). We then measure the response of the system by monitoring at a time $T$ in the future how many cells switched states. We conducted this numerical experiment across a parameter space comparing the size of the stimulus $F_{stim}$ against the out-of-plane tissue stiffness (figure 6 G-H, shows the response to a external stimulus).

Conceptually, the only force that correlates with the height in a positive fashion is the tissue stiffness in the direction orthogonal to the substrate (height) which has the effect of penalizing disagreement between the high and low states. One can imagine a regime in which the tissue elasticity dominates the transition where snap waves propagate, and another regime in low $k_h$ which the switching dynamics is exclusively controlled by a torque mediated stabilization in the current state. 

\section*{Discussion}
\subsection*{Benthic life}
The colonization of marine surfaces was a significant event in the evolutionary history of the animal kingdom\cite{denny_ecological_2016, sperling_placozoan_2010}. Critically, animals did not simply stick to surfaces claiming real estate (as is common of bacterial bio-films or sponges), but began to locomote on surfaces. Ciliary walking provides data bridging the transition between swimming and walking which is achievable without specialized gait control by harnessing the crossover between slipping and stalled. Our minimal model provides a counter-example to the hypothesis that blebbing pseudopodia are the preferred way to move along substrates \cite{noselli_swimming_2019}.

This access to the substrate also enabled new forms of hunting with an external gut by trapping prey underneath the tissue. After prey capture, \textit{T. Adhaerens} releases enzymes into the newly formed cavity and slowly consume its prey\cite{smith2015coordinated, Smith2016AdherensAdhaerens.}. These strategies may have supported ever larger organisms \cite{sperling_placozoan_2010} without the developmental necessity of a gut (formed through developmental process akin to gastulation). Our work, shows that a walking cilium can exploit a bio-mechanical instability to quickly seal the tissue and enable effective hunting. We have shown that this feeding transition is consistent with multiple forms of stimuli including cellular excitability \cite{romanova_sodium_2020, wan_origins_2021} or local neuropeptide release\cite{Senatore2017NeuropeptidergicSynapses., jekely_chemical_2021}. 

\subsection*{Two lenses for excitable mechanics of a walking cilium}

We have shown that the sensory-motor mechanics of a walking cilium is a powerful tool for controlling local and collective behavior. Critically, these rich mechanics shape the mapping of sensory stimulation (coming from the environment) and collective stimulation (initiated by other parts of the organism) onto locomotive forcing. We have shown how the highly nonlinear dynamics of these bio-mechanics can be understood in two primary limits in phase space: (1) high-adhesion, low ciliary tip forcing regime in which ciliary step synchronization is important and (2) a low-adhesion regime in which the phase average can be leveraged to reduce the dimensionality of the dynamics. We have show experimental support for both of these regimes playing an important role in the dynamics of a placozoa's collective locomotion.

Out of the infinite zoo of nonlinear dynamics, our works suggests that both dynamical systems, the ciliary walker and the self-regulating height, can be classified in the general class of excitable dynamical systems which are characterized by large excursions in phase space initiated by a stimulus\cite{fitzhugh_impulses_1961}. The ciliary step can be represented as either a zero or a one \cite{larson_unicellular_2021}. The transition from stepping or not is similar in concept to the excitable dynamics of an integrate and fire neuron, where the probability of taking a step integrates across the force from the other cilia and whether the ciliary beat is 'primed' near the escape phase angle. In a low-dimensional fashion, we can represent this as a $1$ for a step, a $-1$ for a walker which is far from the escape angle and $0$ for a cilium which is primed at the escape phase angle $\psi_{off}$.

\subsection*{Ciliary step excitability}

%From the viewpoint of excitable ciliary steps, the difference in how a ciliary step is triggered becomes clear: the ciliary step is not often triggered by other ciliary steps (e.g. $\rightarrow 1$) but instead by cilia which are generating forces through contact with the surface. The most force comes from neighboring cilia which are primed at $0$ followed by $-1$ with the stepping cilia $1$ generating little-to-no force in the direction of the tissue motion. This integration then adopts a low dimensional relationship which represents the stepping of the ith cilium, $s_i$ as a binary state:
%\[
%s_{i,0\rightarrow 1} = \Theta[\sum_j W_{-1}(h, r) \delta(s_j == -1) + \sum_j W_{0}(h, r) \delta(s_j == 0)]
%\]
%where $\Theta$ is a threshold function of the form of a Heavyside step and coupling can be different for each state of the local cilia.

Practically, step excitability means that steps are more likely when the entire population is on the surface which has the effect of distributing steps over time. If the collective locomotive forcing of a few cells is enough to overcome the stall force then, the population will equally distribute over phase angle. However, if the stall force is greater than the collective ciliary forcing, then the phase density of the population of cilia will accumulate at the stall phase until the population size is large enough to overcome the stall force incurred by the ciliary interaction with the substrate. The evolution of these phase dynamics takes the form of:
$$
\partial_t\psi_i = -\Omega + F_{\psi}(\psi_i) + \sum_j \chi(\psi_j) + \zeta_i
$$
where $\chi$ controls the contribution of a single cilia to the net change of angle of the 

We observe a higher propensity for step synchronization when the tissue is trapped close to the surface due to high local adhesion or internally modulated ciliary forcing. This phenomena was corroborated both numerically and experimentally and demonstrates the critical importance of height in the dynamics of ciliary walking. 

In the supplementary information, we present a simple model which captures the essential character of this form of trapping mediated synchronization with the introduction of three simple parameters, 1) the minimum angular speed $|\omega_{min}(\psi)|$, 2) the collaboration coefficient, $\chi$ which characterizes the leading order contribution to the angular force of another ciliary oscillator tip in contact with the substrate, and 3) the intrinsic noise characteristics of the tissue forcing. In this toy model, we can show the existence of two distinct synchronization transitions. The first of these transitions is controlled by a critical density of trapped cilia which can overcome barrier $\rho_c \sim |\omega_{min}|/\chi$. The second transition is mediated by the mean escape time of a ciliary oscillator driven by a noise process over a barrier relative to the mean return time of the escaped oscillator: $\tau_{escape}/\tau_{return}$ when the return time is long compared to the escape time, the system desynchronizes. 

These excitable and synchronized dynamics are strongly reminiscent of more general phenomena in excitability in depinning dynamics \cite{le_priol_spatial_2021} and stick-slip type oscillations \cite{sugiura_synchronization_2014}.

\subsection*{Height/tissue velocity as an excitable local field potential}

In this work, we show that the emergent mechanics of a walking cilium can be understood analogously to an excitable system whose form is similar to the Fitzhugh-Nagumo equation\cite{fitzhugh_impulses_1961}. This similarity admits a useful analogy where tissue height can be viewed analogously to the resting potential of the individual cell. This height/potential is directly mapped to the local locomotive (time-averaged) forcing causing the tissue to move. From this perspective, our readout of the local height, can be viewed analogously to the spatio-temporal dynamics of a local field potential measurement. 

These measurements show how rapid (correlation time 1 sec) and local changes to the height (correlation length 100 um) are common in the tissue dynamics as the organism locomotes across the flat substrate. These measurements coupled with the local field analogy evokes connections to bioelectric domains of spatially coupled cells\cite{mcnamara_bioelectrical_2020}. 

Going one step further in this simple animal, we have shown that there is a non-linear relationship between the local height and the amount of force the cilium generates to move the tissue. By continuing this analogy, this nonlinear map takes the place of the neuromuscular synaptic network which links the internal cellular potential as a point in the brain to the muscular actuation. Thus by mapping height onto activity, we can study the direct consequences of height fluctuations (or changes to the local cell 'potential') have on the behavior of the moving animal which can be complicated by embodied mechanics of the muscle itself \cite{wang_neuron_2020}.

\subsection*{Understanding our models by analogy to the temporal and rate coding hypothesizes}

The two major classes of models presented in this paper can be viewed as two hypothesizes for what is the minimal representation of internal cell state. There is a parallel conversation which has a long history in the study of neurons \cite{cobb_idea_2020}. The full spectrum of models ranging from neurons as binary switches to the Hodgkin Huxley model have found important applications in the science of the brain \cite{dayan_theoretical_2001} depending upon the prevailing hypothesis of what features matter for a particular phenomena. We suggest that these two modeling paradigms can be classified by analogy to two hypothesis in the study of neurons in how the information is carried: the temporal code and the rate code \cite{dayan_theoretical_2001}. 

The temporal code states that the precise timing of the action potentials in the nervous system carry essential information \cite{dayan_theoretical_2001, sterling_principles_2015}. By analogy, the stepping models presented in this work keep track of the instantaneous (low dimensional) state of the cilia as it takes a step recognizing that the details of the phase angle and amplitude have important ramifications for the mapping onto propulsive force. In this way, this class of model is conceptually analogous to spiking models.

In contrast, the rate code suggests that the relevant information is carried in their frequency of firing, and not in the temporal timing of the action potentials \cite{dayan_theoretical_2001}. This approach is one of the simplest ways to represent continuous numbers with binary variables. By analogy, the phase averaged dynamics in which we couple height to its own dynamical system captures a similar sentiment. Here, it is not the precise timing of the steps, but instead the local tissue height which controls the expected value of the forcing. We leverage the desynchronization (experimental, analytical and within the phase-amplitude-curvature model) at high tissue heights to derive an emergent height-ciliary forcing relationship. 

When a height sensitivity parameter exceeds a critical value, the height dynamics undergo a saddle-node bifurcation -- unfolding a single fixed point into three through the creation of a stable and unstable fixed point pair. The central unstable fixed point defines a separatrix around which a pitch-velocity instability stabilizes either a high tissue height or a low tissue height. While these equations are conceptually very similar to the Fitzhugh-Nagumo equations, they differ in a conceptually critical way with important ramifications for the resulting dynamics. The fundamental coupling between the horizontal nullcline and the vertical nullcline means that the dynamics of single walker emergent dynamical system can not undergo a Hopf bifurcation (the stable fixed point becoming a limit cycle) and can only be a mono-stable or bistable system. 

The bistability observed in the single-walker can be combined with a slow updating variable (the tissue velocity) to generate an excitable type dynamics where the slow variable can be thought of as a the delayed response of the rest of the tissue. Since a reduction in the tissue velocity has the effect of destabilizing the low height fixed point, the tissue velocity plays the role of a slower updating recovery variable which can encourage a return to the high-state in time. In this way, height bistability can give rise to an effective excitable dynamics.

\subsection*{Excitable mechanics complement chemical excitability}

We encourage the reader's skepticism that excitable mechanics is the whole story governing the richness of placozoan behavior, it is not. This paper represents the first in a three paper series on ciliary flocking (how an organism can coordinate ciliary activity without a central controller leveraging mechanics). Part 2 dives into the implications of a rotational degree of freedom and Part 3 explores how emergent quasi-particles shape the underlying collective manifold. This work fits into a broader paradigm on a classification of these stable yet sensitive dynamics into an organism centered perception-action cycle. From this perspective, we propose a mutually-inclusive classification program analogous to Braitenberg's synthetic psychology\cite{braitenberg_vehicles_1986} and explore how existing techniques help to classify experimentally/theoretically probable high-dimensional, sufficiently-nonlinear, coupled dynamical systems \cite{tanaka_recent_2019}. 

There is still much work to be done understanding how ciliary mechanical excitability interacts with the upper epithelium's cortical excitability \cite{armon_epithelial_2020} and a more deep relationship to small signaling molecules\cite{jekely_chemical_2021, Senatore2017NeuropeptidergicSynapses., romanova_sodium_2020}. We look forward to the lessons this 'simple' animal can teach us in how these systems can be wired together. Some of the most compelling candidates are: (1) chemical control of ciliary beat force and direction, (2) chemical and environmental control of surface adhesion. The coupling of these dynamics with chemical control remains a promising direction with much to teach us.

\subsection*{The role of single cell/organelle computation}
And yet, survival requires being not only in-tune with the other cells in the organism but also with the environment. By communication (mechanical, chemical), the collection of cells becomes an animal stable against noise but sensitive to signal via mechanisms analgous to quorum sensing\cite{mathijssen_collective_2019, ellison_cellcell_2016, mukherjee_bacterial_2019}. 

En-route to the collective dance, our understanding must begin at the level of the single cell. In this work, we followed the tradition of developing models by construction and carefully distilling them into more general lessons. An undoubtedly enriching thread would continue this line of inquiry along the parallel thread of viewing the sensory-action map of the neuromuscular system as a form of computation \cite{arcas_what_2000, wang_neuron_2020}. Recent growth following this line of thought for single neurons have shown that the individual is capable of mathematical operations previously thought to be only possible in ensembles \cite{li_dendritic_2019, poirazi_pyramidal_2003, gidon_dendritic_2020}. 

The extension of this line of inquiry to cilia as a rudimentary yet rich alternative to the neuromuscular system has important implications \cite{gilpin_multiscale_2020, wan2018time, wan_origins_2021, smith_coherent_2019, Senatore2017NeuropeptidergicSynapses., pfeifer2007self} in understanding the general principles of how the animal kingdom reconciles stability with sensitivity. As we more deeply appreciate the diversity of rapid sensory transduction techniques, we can take a step back to appreciate the universal aspects and unifying principles which arise from diverse and complicated constructions. 

\section*{Conclusion}

In this work, we set out to understand how a simple animal moved along marine surfaces. Our direct experimental measurements of the rich ciliary dynamics of 'walking' on a substrate lead us to ask: is excitable ciliary walking sufficient to explain the collective dynamics of the organism? In developing a model by construction in the form of a phase-amplitude-curvature oscillator interacting with a sticky substrate, we discovered two forms of excitable dynamics at the core of ciliary walking as a tool of locomotion. The first suggests that ciliary steps can be viewed through the lens as an excitable system and the second identified a mechanical analog to the local field potential in the spatio-temporal height field. Linking ciliary forcing to a dynamical system in height also reveals a bistable system which can accomplish the critical tasks of cell-to-cell communication and environmental sensitivity. These results reaffirm the utility of this ciliary toolkit for rapid and versatile sensory-action loops in an animal without a nervous system.

\section{Acknowledgements}
We thank all members of the PrakashLab for scientific discussions and comments. In particular, we thank Pranav Vyas, Shahaf Armon, Grace Zhong, Hazel Soto-Montoya and Vivek Prakash in the lab for their contributions to a vibrant research community. M.S.B. and L.K. gratefully acknowledge support by the National Science Foundation Graduate Research Fellowship (DGE-1147470) and the Stanford University BioX Fellows Program (M.S.B). This work was supported by HHMI Faculty Fellows Award (M.P), BioHub Investigator Fellowship (M.P), Pew Fellowship (M.P), Schmidt Futures Fellowship, NSF Career Award (M.P), NSF CCC (DBI-1548297) and the Moore Foundation. 

\bibliography{mybib.bib}

\section{Figures}
\onecolumngrid
\begin{figure}
\includegraphics[width = 0.9\textwidth]{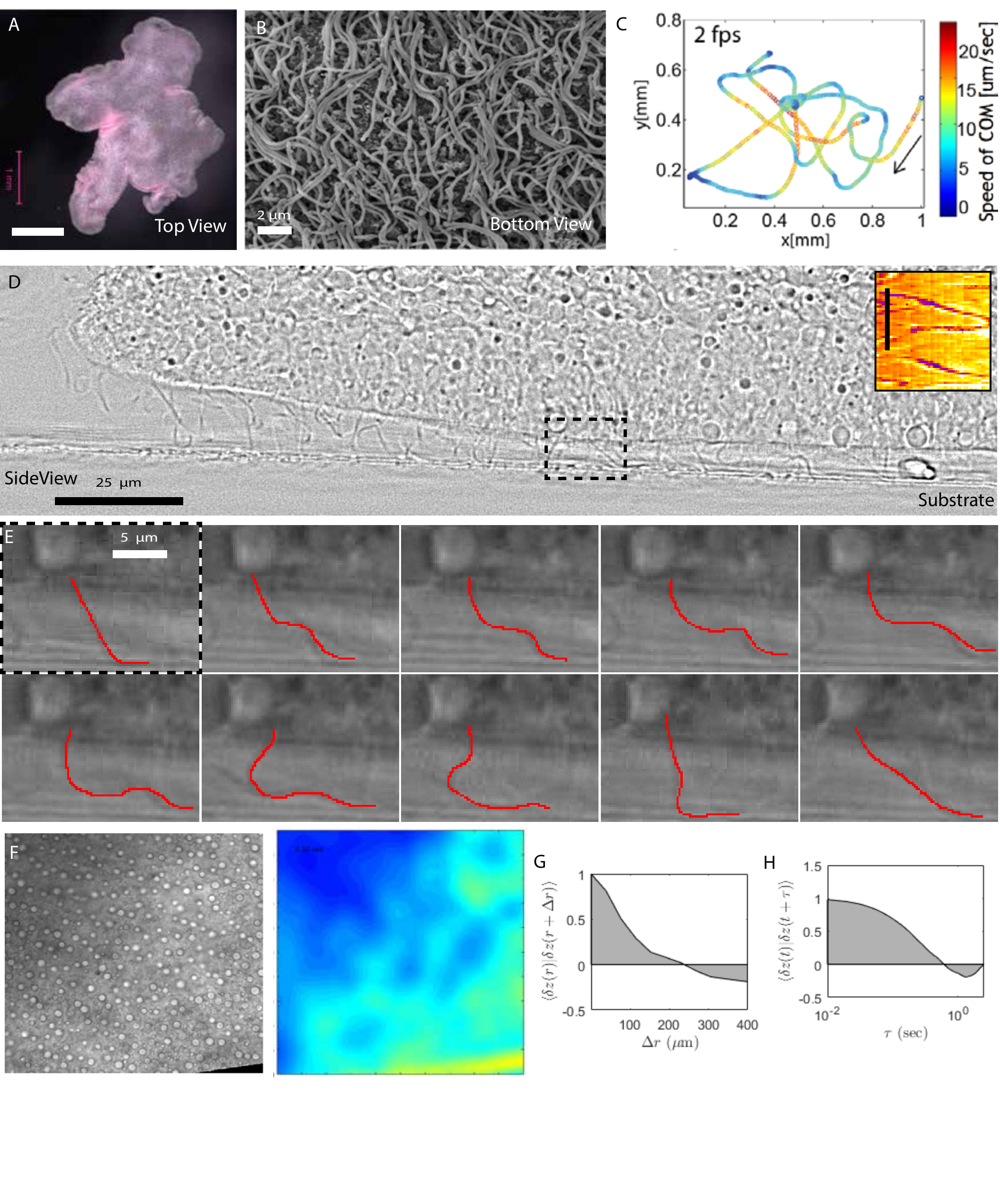}
%\caption{\textbf{FIG 1}: (Caption next page.)}
  \label{fig:fig1}
\end{figure}

\begin{figure}
\caption{FIG 1: \textbf{Placozoa use cilia to walk in a height-dependent manner.} A) The three layer body plan of placozoa flatten to move on substrates. B) The bottom layer of the organism is comprised of high density mono-ciliated cells captured by Scanning Electron microscopy. C) The organism exhibits dynamic locomotion influenced strongly by environmental variability. D) High-speed microscopy from the side view of the organism demonstrates strong ciliary interaction with the substrate deviating cilia shapes from periodic beats in a resistive fluid. E) Sampled at 10Hz, a walking cilia sweeps out a ‘step’ consistent with a traveling wave from cell to ciliary tip.  F) The spatio-temporal dynamics of tissue height can be extracted using a calibrated measure of image sharpness and reveal rapid height fluctuations with G) correlation lengths on the order of ~200 $\mu$m in space and H) seconds in time.}
\end{figure}

\newpage

\begin{figure}
\includegraphics[width = \textwidth]{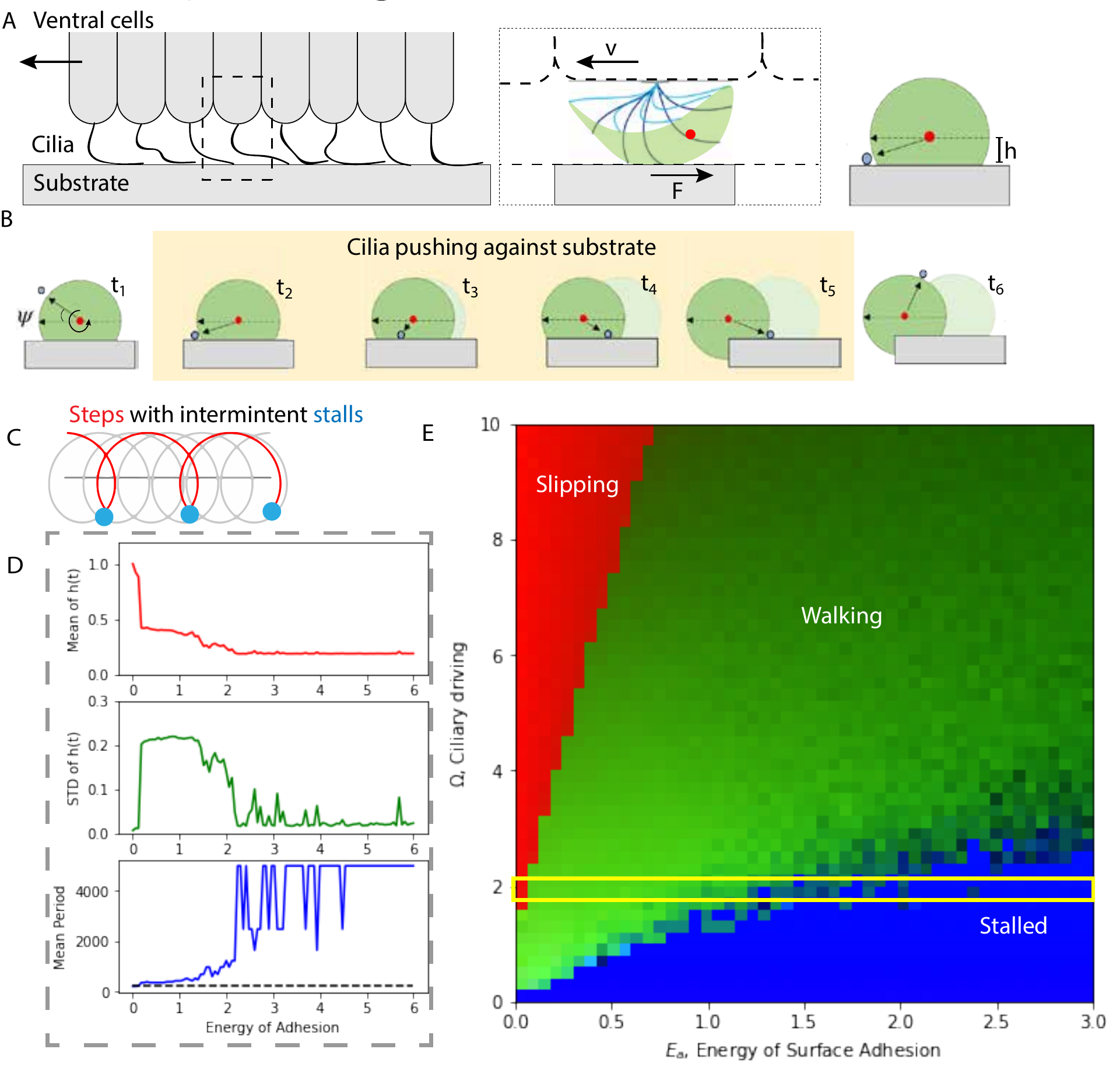}
\caption{\textbf{FIG 2}: (Caption next page.)}
  \label{fig:fig2}
\end{figure}

\begin{figure} [t!]
\caption{FIG 2 (previous page) \textbf{A model demonstrates that the self-stabilizing ciliary gait is sensitive to both environmental and internal states.} A) We low dimensional our understanding of a walking cilium by developing a model which reflects the dynamics of ciliary oscillation in conjunction with substrate interaction. B) In the simplest model which captures these essential ingredients, we merge a successful model in the literature with substrate interaction -- the view point can be represented as a ciliary tip trajectory competing with substrate interaction. C) The motion of both the tissue and the ciliary tip couple to  D) generate cyclotron-type tip orbits when in contact with the surface, red, punctuated by intermittent stalling in blue. E) We conduct extensive numerical studies of this single-cilia oscillator as a function of both the substrate adhesion energy and the ciliary forcing. F) A two dimensional phase space coloring each pixel by the three summary statistics on each channel, shows a crossover between ciliary slipping, red, (where the cilia pushes off the substrate and interacts very little -- analogous to a tire without traction), a walking region (green) and a stalled regime (blue). These uncontrolled ciliary gaits organize in a fashion which is sensitive to both the internal ciliary state and the environmental conditions. }
\end{figure}

\newpage
\begin{figure}
\includegraphics[width = \textwidth]{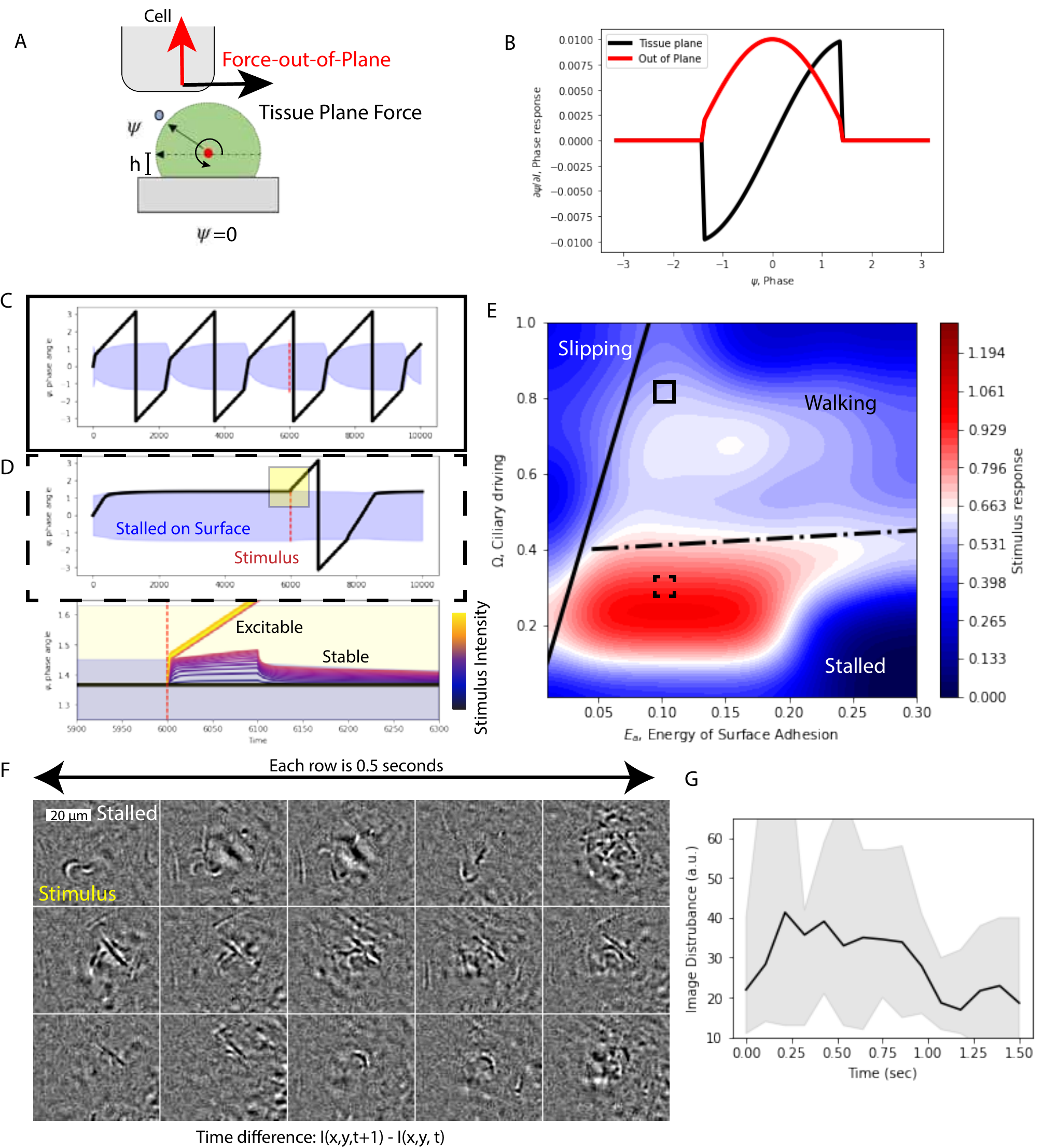}
\caption{\textbf{FIG 3}: (Caption next page.)}
  \label{fig:fig3}
\end{figure}

\begin{figure} [t!]
\caption{FIG 3 (previous page) \textbf{ A stalled cilium can be excited into taking a step by a stimulus.} A) In the phase-amplitude oscillator depiction, there are two orthogonal directions to apply force, in the tissue plane and out of the tissue plane.  B)The phase response curves for both in-plane (black) and out-of-plane forces (red) look like sines and cosines respectively, which are truncated when the cilium pulls away from the surface. C) We can plot the evolution of the phase angle (black) and the time evolution of the of the subtending surface angle (blue). At a predefined moment, we subject the system to a stimulus and measure the difference between the resulting trajectory and the undistributed trajectory $\psi(t | I\neq 0) - \psi(t | I=0)$. In the case seen in part C, the cilia is oscillating at a frequency which is modified by the interaction with the surface, but is not influenced by the stimulus. D) Conducting the same numerical experiment at lower ciliary driving force, the cilia stalls until the stimulus pushes it off the surface where it subsequently continues it surface-free trajectory. In the zoom region (yellow) we compare the response of the system for different amplitudes of stimulus and show that for sufficiently strong stimuli the ciliary phase angle behaves like a excitable system where the stimulus can initiate a step. E) We conduct this numerical experiment across a wide region of parameter space spanning the slipping-walking-stalling regimes of the single cilium and show that system responds the most along the boundary between stalled and walking. F) In experiments where the tissue begins in a stalled state (low height, tissue velocity close to zero), we observe the avalanche type excitability of ciliary steps in response to a stimulus. G) We summarize the adjacent timeseries snapshots with the mean amplitude of the image disturbance (with Gaussian smoothing) over time, the mean and extremes are represented by the line and fill region respectively. This timeseries shows a large local response which relaxes back to background over $\sim 1$ sec and $\sim 8$ ciliary steps.}
\end{figure}

\newpage
\begin{figure}
\includegraphics[width = \textwidth]{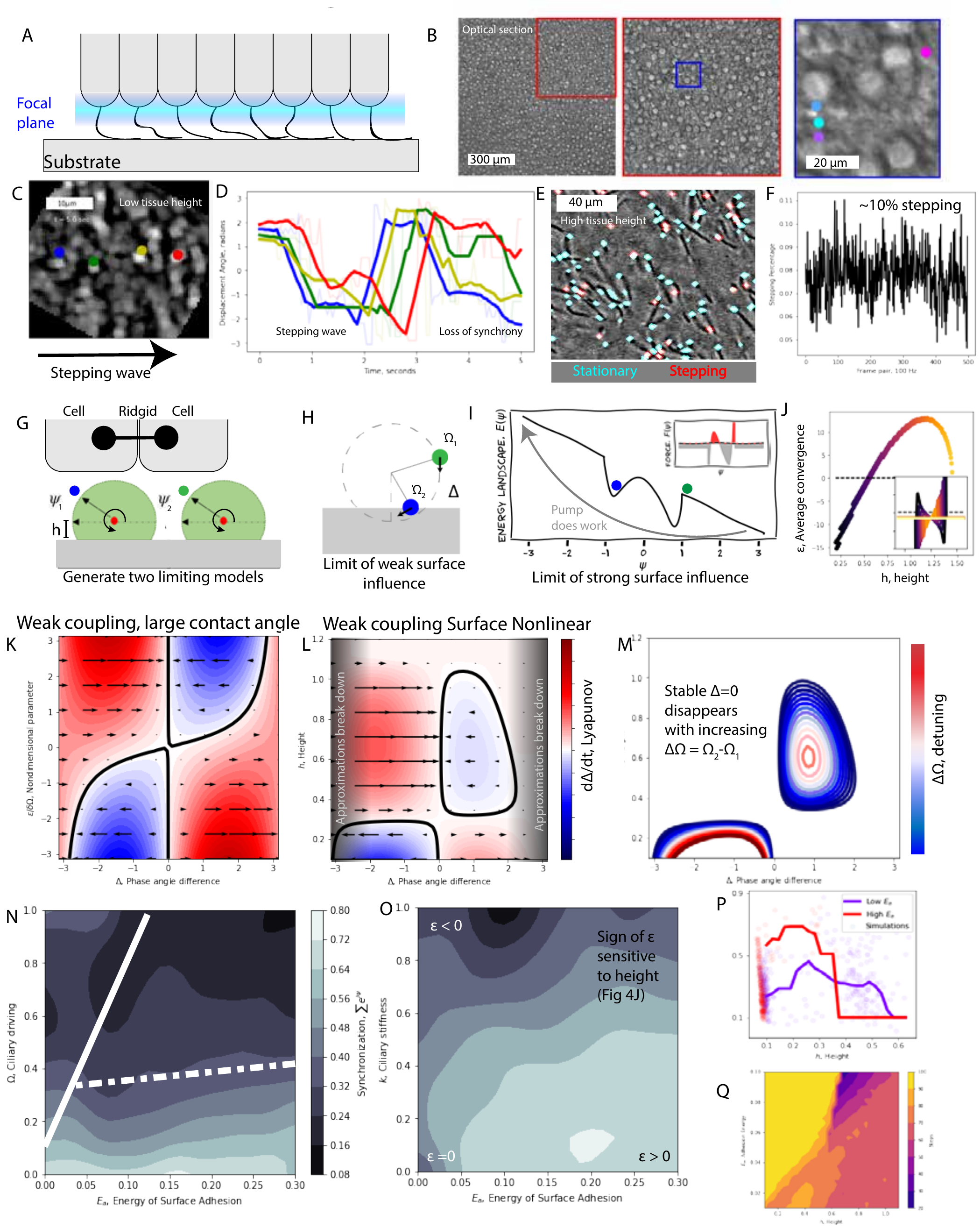}
\caption{\textbf{FIG 4}: (Caption next page.)}
  \label{fig:fig4}
\end{figure}

\begin{figure} [t!]
\caption{FIG 4 (previous page) \textbf{Ciliary oscillators interacting through the tissue and substrate weakly synchronize at an intermediate height.} A) Using single-organelle resolved microscopy (Nikon 60x-TIRF) at high speed, we focus our objective in the ciliary cross-section plane, where individual ciliary crossections show up as $\sim 20 $ pixel dark occlusions of the structured tissue background. B) We remove the tissue spatial frequencies by band-pass filtering and traverse through the scales to find regions of the tissue which are at different relative heights (as measured by our sharpness technique in fig 1).  C) We computationally register 4 cilia in a line at low height and study their beating as a step wave passes across them. D) Plotting up a proxy for the phase angle of the individual cilia shows that the first step in the movie exhibits a passing wave from left (blue, green, yellow) to right (red).  In the second beat observed, the tissue has pulled down closer to the surface and the step synchrony is lost.  E) At a higher relative height, we find no step synchrony across the field a view and plot up the spatial distribution of stepping cilia (red) on top of segmented cilia (teal).  F) The timeseries of the stepping percentage shows a uniformity in ciliary step density with about $\sim 8- 10 \%$ stepping at any one time with only small fluctuations indicative of an approximately uniform ciliary phase angle density.  H) We can understand the accessible dynamics of pairs of ciliary oscillators connected via a rigid tissue. I) We present two limiting models, the first in the limit of weak surface influence (e.g. ciliary driving is strong compared to surface forces in the slipping regime from figure 2E). J) The second limiting case takes the form of ciliary coupling small compared to the forces coming from the substrate which is consistent with the stalling regime. K) We define a parameter which is helpful in the analysis, how much ciliary density increases or decreases over a ciliary contact. We call this quantity $\epsilon$, the average ciliary density convergence, and it is calculated numerically as a function of surface height, $h$ for a fixed energy of adhesion $E_a$ and oscillator amplitude stiffness $k$ to show a function which crosses zero with a positive slope at a critical height, before slowly decaying back toward zero at the height above which the cilium no longer sticks to the substrate. L) For the two-oscillator, weak surface interaction case, we can plot the result of the Adler-like equation under the approximation that a rotating reference frame holds (e.g. a large contact angle) to show the Lyapunov exponent of the phase angle difference between the two oscillators for every phase angle and choice of non-dimensional parameter average convergence over natural frequency detuning: $\epsilon/\Delta\Omega$. This plot reveals a crossover from a stable $\Delta = 0$ fixed point to an unstable $\Delta =0$ fixed point with the changing sign of the average convergence $\epsilon$.  ...(continued on next page) ...}
\end{figure}
\begin{figure}[t!]
\caption{FIG 4 ...(continued from previous page) ...
M) Changing mathematical technology to relax the rotating reference frame assumption and replacing $\epsilon$ with its function as height, we find that the fixed point at $\Delta = 0$ is stable at intermediate heights for small native frequency detuning, $\Delta \Omega$. N) We also show that the height range of this stable fixed point is sensitive to native frequency detuning, $\Delta\Omega$ and the fixed point disappears entirely at a critical detuning amplitude.  O) We corroborate these results with a full numerical model of N=100 ciliary oscillators coupled all-to-all for a sweep of parameters through the ciliary forcing $\Omega$ and energy of adhesion $E_a$ plane to show that synchronization is largest in the stalling regime. P) Studying the all-to-all model in the energy of adhesion $E_a$ versus amplitude-direction stiffness $k$ plane supports the role of the average compression $\epsilon$ on synchronization (see SI for $\epsilon$ plotted over the same regime). Q) We can summarise the $N \sim 10^3$ simulations by looking at the emergent relationship between synchronization and height which suggests that there is an intermediate height at which synchronization occurs and if the tissue is too high or too low, the synchronization is suppressed. R) To tie these results back to experiment, we show that a $\sim 10\%$ stepping with uniform phase angle density is achievable at an intermediate height (above stall) yet significant energy of adhesion.  These results suggest that low step frequencies may be close to the walking-stalling transition.
}
\end{figure}

\newpage
\begin{figure}
\includegraphics[width = \textwidth]{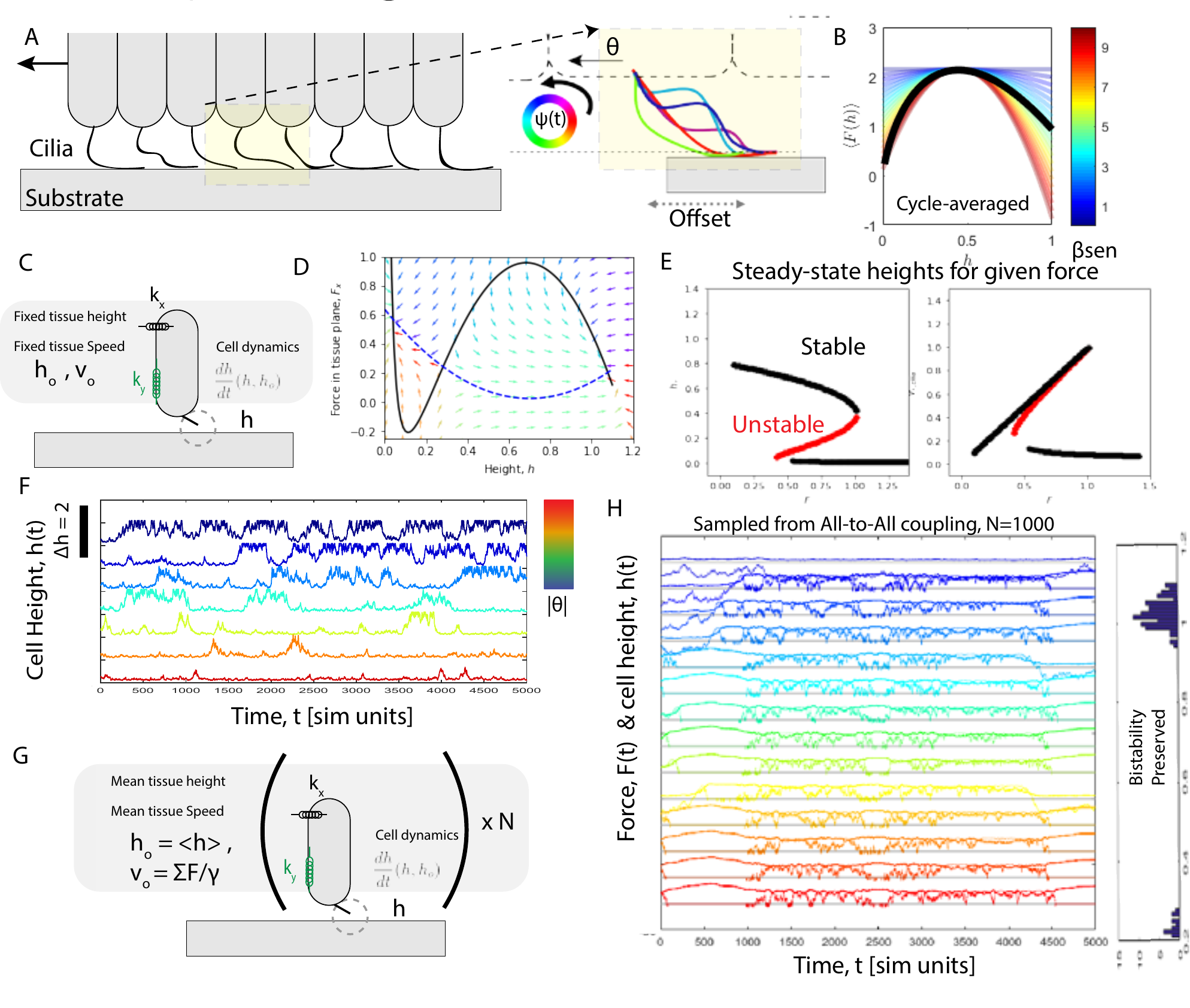}
\caption{\textbf{FIG 5}: (Caption next page.)}
  \label{fig:fig5}
\end{figure}

\begin{figure} [t!]
\caption{FIG 5 (previous page) \textbf{The central importance of height-forcing coupling as a generator of spatio-temporal bistability in walking tissues.} A) Ciliary walking sweeps out cyclic trajectories in shape space which can be modified by substrate interaction. B) The cycle averaged force generatation as a function of height exhibits a biphasic (slope increasing and decreasing around a maximum) character well captured by a second order Taylor expansion. C) The dynamics of the self-regulating height is analogous to the a wheelie mechanism on a bike -- ciliary force applied to the substrate results in a torque which counteracts the force applied to the tissue. When the two forces match (e.g. the force coming from the tissue and the force generated by the walking cilia’s substrate interaction), the torque balances and the height regulates. D) The interplay between in plane tissue force and height creates a 2d dynamical system with a rich set of null-clines, which intersect for two stable fixed points and a third unstable fixed point. E) Long time integration of the governing equations (black) and the negative of the governing equations (red) shows the bistable asymptotic behavior of the height and locomotive forcing for different values of externally applied force along the x axis. F) Numerical time courses of the equations integrated via Euler-Maruyama method (dt = $10^{-4})$ for various choices of the externally imposed tissue force shows the crossover between preferentially occupying the high state at low force and the low state at high force. Tissue force can be thought of as a biasing torque causing the pitch angle to nose-dive. G) To confirm if the bistability is preserved under mean field closure of the tissue force (rather than an externally imposed parameter), we study an all-to-all coupled model of these dynamical systems. H) For N=1000, the plot of a randomly selected subset shows both examples of cells which push to the high-state and examples in the low state at long times, following long lived transients. We histogram the final height after long simulations to show that bistability appears to be conserved under this form of closure.} 
\end{figure}

\newpage
\begin{figure}
\includegraphics[width = 0.9\textwidth]{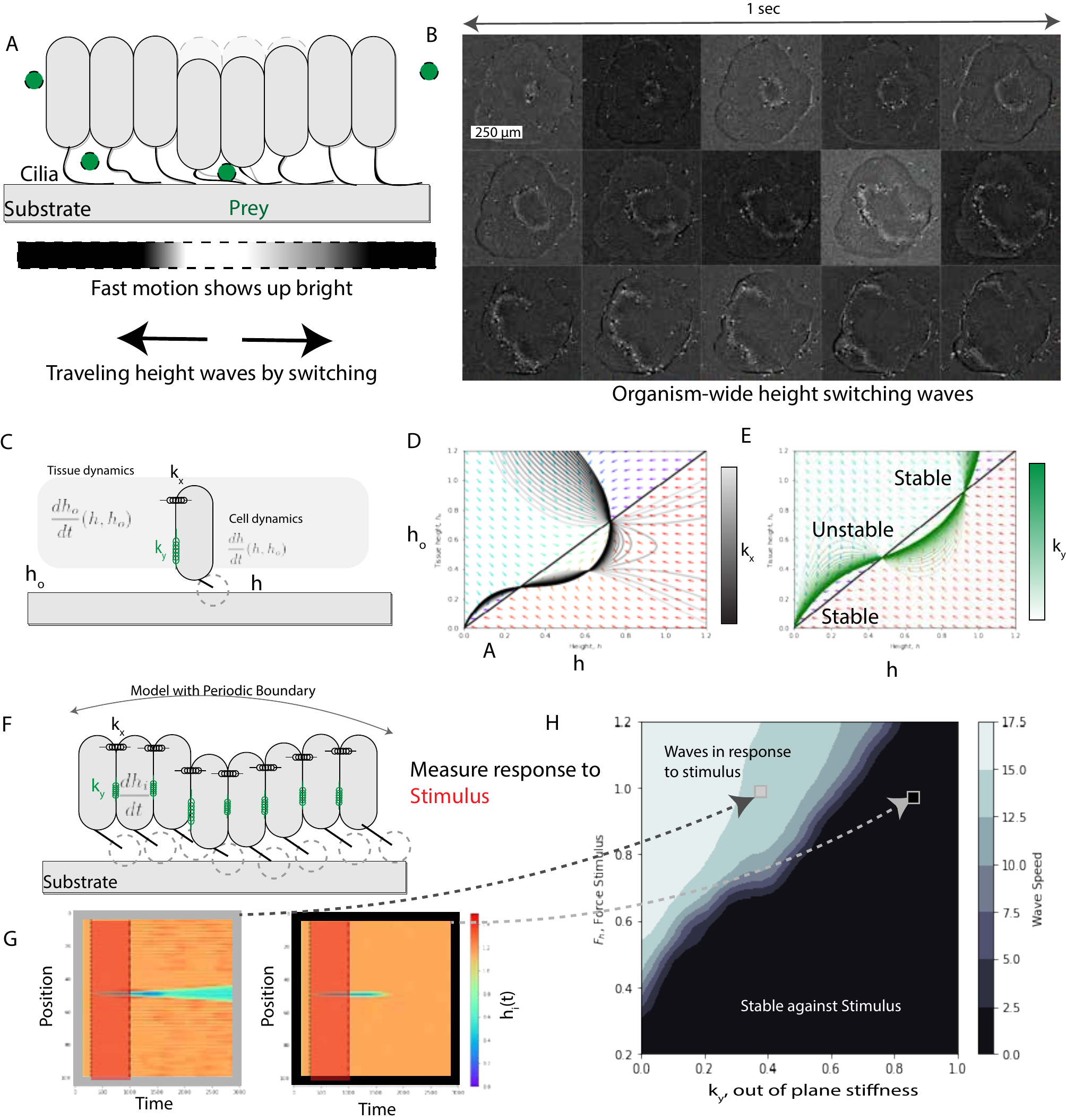}
\caption{\textbf{FIG 6}: (Caption next page.)}
  \label{fig:fig6}
\end{figure}

\begin{figure} [t!]
\caption{FIG 6 (previous page) \textbf{The height bistability can support linear speed traveling waves.} A) A placozoan in the presence of food will periodically press itself against the surface trapping algae-- a behavior associated with feeding with an external gut. We use the image disturbance field $I(x,y,t+1) - I(x,y,t)$ to show regions with fast motion as bright.  B) Visualizing the image disturbance field in experiment we observe that the transition from high-to-low begins locally and the message is passed on the second timescale across $10^5$ cells via a trigger wave. C) We adapt the cycle-averaged model into a dynamical system relating tissue height $h_o(t)$ to the height of the cell under study, $h(t)$, via leading order elasticity in the tissue plane, $k_x$ and in the height direction $k_y$. D) Using separation of timescales, we plot the two dimensional dynamical system associated with these degrees of freedom for varying $k_x$ to show that the bistability is preserved and that above a critical value of $k_x$ the fixed point at high height begins to oscillate similar to a high integral PI controller. We study the response of the tissue below this critical value. E) The dynamical system in response to increasing $k_y$ causes the null-clines to squeeze in closer to $h_o = h$. F) We study traveling waves by promoting the dynamics to a 1D chain and studying the response of the chain to a stimulus at the center of the chain. G) At low  $k_y$ we observe trigger wave propagation of switching. At high $k_y$ the chain is stable against stimulus pulling the single cell back up to the tissue height. H) The response space shows that small $k_y$ is more susceptible to small stimuli than higher $k_y$.}
\end{figure}

%
% ****** End of file apssamp.tex ******
\end{document}